\documentclass[twocolumn,superscriptaddress,amsfont,amssymb,amsmath, showpacs,balancelastpage, nofootinbib]{revtex4-2}
\usepackage{graphicx,longtable,mathrsfs,color,array}
\usepackage[unicode=true,pdfusetitle,
bookmarks=true,bookmarksnumbered=true,bookmarksopen=true,bookmarksopenlevel=1,
breaklinks=false,pdfborder={0 0 0},backref=false,colorlinks=true]{hyperref}
\hypersetup{citecolor=blue,filecolor=blue,linkcolor=blue,urlcolor=blue,pdfauthor={Name}
}
\usepackage[usenames,dvipsnames]{xcolor}
\usepackage{amssymb,amsmath,mathtools,mathrsfs,enumitem}
\usepackage{epsfig,subfigure,placeins,float}
\usepackage{booktabs,longtable,multirow}
\usepackage{exscale,relsize}
\usepackage[normalem]{ulem}
\usepackage[T1]{fontenc}
\usepackage[utf8]{inputenc}
\usepackage{enumerate}
\usepackage{times, mathptmx}
\usepackage{tikz}
\usepackage{aas_macros}
\usepackage{tabularx}
\usepackage{tabularray}
\usepackage{diagbox}
\usepackage{multirow}
\usepackage{pifont}
    \newcommand{\cmark}{\ding{51}}%
    \newcommand{\xmark}{\ding{55}}%
\newcolumntype{M}[1]{>{\centering\arraybackslash}m{#1}}
\usepackage{hhline}
\usepackage{makecell}
\usepackage{cleveref}
\usepackage{siunitx}
\usepackage{multirow}
\usepackage{booktabs}

\usetikzlibrary{arrows,positioning,decorations.markings,decorations.pathmorphing,calc}

\newcommand{\be}{\begin{equation}}
\newcommand{\ee}{\end{equation}}
\newcommand{\ba}{\begin{align}}
\newcommand{\ea}{\end{align}}

\newcommand{\comment}[1]{}

\newcolumntype{C}[1]{>{\centering\let\newline\\\arraybackslash\hspace{0pt}}m{#1}}
\newcommand*{\backin}{\rotatebox[origin=c]{-180}{$\in$}}%
\newcommand{\mn}{{\mu\nu}}

\newcommand{\nocontentsline}[3]{}
\newcommand{\tocless}[2]{\bgroup\let\addcontentsline=\nocontentsline#1{#2}\egroup}

\def\mpl{M_{\rm Pl}}
\def\Mpl{M_{\rm Pl}}

\newcommand{\red}[1]{\textcolor{black}{ #1}}
\newcommand{\black}[1]{\textcolor{black}{ #1}}

\newcommand{\jnh}[1]{}

\definecolor{hyperref}{RGB}{026,028,087}
\def\gsim{ \lower .75ex \hbox{$\sim$} \llap{\raise .27ex \hbox{$>$}} }
\def\lsim{ \lower .75ex \hbox{$\sim$} \llap{\raise .27ex \hbox{$<$}} }

\def\nn{\nonumber}

\newlength{\stheight}
\newcommand\textst[1][fu-grey]{
\ifmmode\setlength{\stheight}{+1.0ex}
\else\setlength{\stheight}{+0.5ex}
\fi
\bgroup\markoverwith{\textcolor{#1}{\rule[\the\stheight]{2pt}{1.0pt}}}\ULon
} 
 
\newcommand{\textins}[2][fu-grey]{
\ifmmode\mathcolor{#1}{#2}
\else\textcolor{#1}{#2}\@\,
\fi
}
\graphicspath{{./Plots/}}
\allowdisplaybreaks

\tikzstyle{vecArrow} = [thick, decoration={markings,mark=at position
1 with {\arrow[semithick]{open triangle 60}}},
double distance=1.4pt, shorten >= 5.5pt,
preaction = {decorate},
postaction = {draw,line width=1.4pt, white,shorten >= 4.5pt}]

\begin{document}

\setcounter{tocdepth}{5}
\title{Stability and quasinormal modes for black holes with time-dependent scalar hair}

\author{Sergi Sirera}
\affiliation{Institute of Cosmology \& Gravitation, University of Portsmouth, Portsmouth, PO1 3FX, U.K.}
\affiliation{Department of Physics \& Astronomy, University College London, London, WC1E 6BT, U.K}

\author{Johannes Noller}
\affiliation{Department of Physics \& Astronomy, University College London, London, WC1E 6BT, U.K}
\affiliation{Institute of Cosmology \& Gravitation, University of Portsmouth, Portsmouth, PO1 3FX, U.K.}

\begin{abstract}
We investigate black hole solutions with time-dependent (scalar) hair in scalar-tensor theories. Known exact solutions exist for such theories at the background level, where the metric takes on a standard GR form (e.g. Schwarzschild-de Sitter), but these solutions are generically plagued by instabilities. Recently, a new such solution was identified in \cite{Bakopoulos:2023fmv}, in which the time-dependent scalar background profile is qualitatively different from previous known exact solutions - specifically, the canonical kinetic term for the background scalar $X$ is not constant in this solution. We investigate the stability of this new solution by analysing odd parity perturbations, identifying a bound placed by stability and the resulting surviving parameter space. We extract the quasinormal mode spectrum predicted by the theory, identifying a shift of quasinormal mode frequencies and damping times compared to GR. We forecast constraints on these shifts (and the single effective parameter $\hat\beta$ controlling them) from current and future gravitational wave experiments, finding constraints at up to the ${\cal O}(10^{-2})$ and ${\cal O}(\red{10^{-4}})$ level for LVK and LISA/TianQin, respectively.
All calculations performed in this paper are reproducible via a companion Mathematica notebook \cite{ringdown-calculations}.
\end{abstract}

\date{\today}
\maketitle

\tableofcontents

\vfill\null

\section{Introduction} \label{sec-intro}

{\bf Ringdown tests of gravity}: The rise of gravitational wave science offers a new way to probe gravity in the strong field regime. In particular, the final stage of gravitational wave signals emitted by binary compact object mergers, known as the ringdown, can be notably sensitive to new gravitational physics. The ringdown phase is well modelled by linear perturbations on a black hole background, whose evolution is described by a superposition of complex decaying frequencies, or quasinormal modes (QNMs) \cite{Berti:2009kk}. General Relativity (GR) predicts that the full set of QNMs is fixed by the black hole's mass and angular momentum (and charge, if present). This is a consequence of no-hair theorems in GR \cite{Mazur:2000pn}, and is therefore generically violated in extensions of GR which allow for hairy solutions.\footnote{\red{Note, however, that the presence of hair is not a necessary requirement to violate the GR spectrum. In particular, the equations governing perturbations (hence QNMs) can differ from those of GR, even in cases where the background solutions are those of GR without any additional hair 
\cite{Barausse:2008xv}. For example, even parity QNMs can be modified in Horndeski scalar-tensor theories in the absence of non-trivial hair \cite{Tattersall:QNMsHorn}.}} In such cases, QNMs can depend on extra parameters associated with the hair, and so looking for deviations from the GR-predicted QNM spectrum provides a powerful null test of GR. More quantitatively, QNM measurements can therefore be used to place constraints on (additional) fundamental gravitational degrees of freedom generically associated with extensions of GR that leave an imprint on the QNMs \cite{Chen:2021cts}. This program, sometimes referred to as testing the \textit{Kerr hypothesis}\footnote{Formally, the Kerr hypothesis asks whether the final stage after gravitational collapse is well described by the Kerr geometry \red{in GR at the background and perturbative level}, and therefore contains no hair \cite{Bambi:2011mj}. \red{As mentioned in the previous footnote, note that the absence of hair is not a sufficient condition to prove the Kerr hypothesis.}}, or \textit{black hole spectroscopy} \cite{Dreyer:2003bv}, has been employed to probe the strong gravity regime in a diverse number of ways, see e.g. \cite{Lahoz:2023csk,Bamber:2022pbs,Taylor:2024duw,Tattersall:BHspectro,Tattersall:QNMsHorn,Pitte:2024zbi,Berti:2005ys,Cardoso:2019rvt,Carullo:2019flw,Meidam:2014jpa,Brito:2018rfr,Cardoso:2019mqo,Saketh:2024ojw,Mukohyama:2023xyf,Pierini:2021jxd,Wagle:2021tam,Brito:2018hjh}.
\\

{\bf Black hole-scalar solutions}:
In this paper we will work in the context of scalar tensor theories, more specifically Horndeski (scalar-tensor) gravity \cite{Horndeski:1974wa,Deffayet:2011gz}, which is the most general theory built with a metric tensor and a scalar field yielding second-order equations of motion. Black hole solutions to such theories broadly divide into two categories. On one hand, we have \textit{stealth} black hole solutions, where the background metric takes the same form as known GR solution such as Schwarzschild or Schwarzschild-de Sitter (SdS). 
Any potential `hair' associated to such stealth solutions is then associated only with the profile of the background scalar.
On the other hand, due to the increased complexity of scalar tensor actions with respect to the Einstein-Hilbert action, more general black hole solutions can also exist in such theories. Thus, the other category of black hole solutions involve new background solutions for the metric itself (as well as for the scalar) -- see e.g. \cite{Anson:2020trg,Babichev:2023mgk}. In this paper, we focus on the former, i.e. stealth black hole solutions with non-trivial scalar profiles. In particular, we will focus on Schwarzschild and 
SdS black hole solutions.
Especially SdS solutions have been a focus of attention in previous studies \cite{Babichev:2013cya,Kobayashi:2014eva,Babichev:2012re,Babichev:2016fbg,Babichev:2017guv,Babichev:2016kdt,BenAchour:2018dap,Motohashi:2019sen,Takahashi:2020hso,Takahashi:2021bml,Bakopoulos:2023fmv,Babichev:2016rlq}, 
since the de Sitter asymptotics at large distances more closely mimic realistic black holes embedded in cosmological space-times (in comparison to Schwarzschild solutions with their Minkowski asymptotics).   
In section \ref{sec-CBH} we will recap S(dS) solutions found in scalar-tensor theories, where the scalar has a time-dependent profile. But looking ahead, known exact solutions 1) generically suffer from instabilities, and 2) have been investigated for scalar profiles where $X\equiv-\frac{1}{2}\phi_\mu\phi^\mu$, the kinetic term for the scalar $\phi$, is constant.

In this paper we therefore focus on a novel black hole solution that was recently found within Horndeski scalar-tensor theories \cite{Bakopoulos:2023fmv} and is a promising candidate going beyond the constant $X$ assumption and possibly providing stable dynamics as well. This theory is given by the following Lagrangian 
\begin{align}
    {\cal L}= 2\eta\sqrt{X} -2\Lambda + R(1+\lambda\sqrt{X}) + \frac{\lambda}{2\sqrt{X}}\left[(\square\phi)^2-\phi^{\mu\nu}\phi_{\mu\nu}\right],
    \label{eq-theory}
\end{align}
where one can see that standard GR plus a decoupled scalar (albeit with non-standard kinetic term) is recovered in the $\lambda \to 0$ limit. This theory possesses an exact background solution of the form \cite{Bakopoulos:2023fmv} 
\begin{align}
ds^2&=-B(r)dt^2+\frac{1}{B(r)}dr^2+d\Omega^2, \nn \\
\bar\phi & = qt+\psi(r),
\label{backg}
\end{align}
with
\begin{equation}
    B(r)= 1-\frac{2M}{r}-\frac{1}{3}\Lambda r^2.
    \label{eq-BSdS}
\end{equation}
The background solution for the metric is therefore manifestly of SdS form and $q$ is a (dimensionful) constant characterising the linear time-dependence of the scalar field, with $\psi$ encoding its radial dependence. As we are dealing with a shift symmetric theory, the derivatives of the scalar field encode its most important features. In particular, the radial derivative of the scalar field and the kinetic term $X$ are given by
\begin{align}
    \psi'^2&=\frac{q^2}{B^2}\left(1-\frac{\lambda B}{\lambda+\eta r^2}\right), \nn \\
    X &=\frac{1}{2}\frac{q^2\lambda}{\lambda+\eta r^2},    
    \label{eq-psiprofile}
\end{align}
where it is immediately apparent that this is a scalar profile with non-constant $X$ and we illustrate the dependence of $X$ on $r$ in Figure \ref{fig:X-plot}.
Our focus in this paper will therefore be to investigate this novel solution further, specifically its stability and QNM spectrum.
\\

{\bf Outline}:
The paper is organised as follows. In section \ref{sec-CBH} we recap relevant hairy black hole solutions -- in particular stealth black hole SdS solutions with linearly time-dependent scalar profiles -- discovered so far and discuss related stability issues. In section \ref{sec-bh-stab-perts} we examine the main theory of interest for this paper \eqref{eq-theory}
at the background and perturbative level.
We investigate the behaviour of linear odd parity perturbations on the hairy black hole background \eqref{backg} and show when such perturbations are stable. In section \ref{sec-qnms} we derive the corresponding modified Regge-Wheeler equation and employ the WKB method to extract the associated QNM spectrum. Finally, in section \ref{sec:forecasts} we forecast constraints on the parameter that controls deviations from GR via a Fisher information analysis, before concluding in section \ref{sec:conclusions} and collecting further relevant details in the appendices. Throughout we work in \textit{geometric} units, where $c=\hbar=G=1$, unless explicitly stated otherwise.

\section{Scalar-tensor theories and hairy black holes}
\label{sec-CBH}
In this section we briefly review the current state of hairy black hole solutions and their known stability properties.
Our focus is on theories where the new fundamental physics is characterised by a single scalar degree of freedom $\phi$, thus constituting a scalar-tensor theory. As previously noted, Horndeski gravity is the most general such theory resulting in second-order equations of motion \cite{Horndeski:1974wa,Deffayet:2011gz}\footnote{For the equivalence between the formulations of \cite{Horndeski:1974wa} and \cite{Deffayet:2011gz}, see \cite{Kobayashi:2011nu}.}, and is governed by the following action
\begin{align}
    S= \int &d^4x \sqrt{-g}\Big[G_2 + G_3 \Box \phi
    + G_4 R + \nn \\
    &G_{4X}\left[(\square\phi)^2-\phi^{\mu\nu}\phi_{\mu\nu}\right] +G_5G_{\mu\nu}\phi^{\mu\nu}- \nn \\ &\frac{1}{6}G_{5X}\left[(\square\phi)^3 -3\phi^{\mu\nu}\phi_{\mu\nu}\square\phi+2\phi_{\mu\nu}\phi^{\mu\sigma}\phi_\sigma^\nu\right]\Big].
    \label{S}
\end{align}
Here we have introduced the short-hands $\phi_\mu\equiv\nabla_\mu\phi$ and $\phi_{\mu\nu}\equiv\nabla_\nu\nabla_\mu\phi$, the $G_i$ are free functions of $\phi$ and $X$, where we recall that $X\equiv-\frac{1}{2}\phi_\mu\phi^\mu$, and $\square$ is the d'Alembert operator. When discussing previous work on black hole solutions in scalar tensor theories, we will occasionally also refer to solutions derived in extensions of Horndeski theories, in particular within DHOST \cite{Langlois:2015cwa,Crisostomi:2016czh,BenAchour:2016fzp,Takahashi:2017pje,Langlois:2018jdg}. 

Horndeski gravity admits a richer variety of black hole solutions compared to GR. Yet, akin to GR, there exist a number of no-hair theorems for a wide range of scalar-tensor theories which enforce the scalar to possess a trivial profile \cite{hawking1972black,PhysRevLett.108.081103,PhysRevLett.110.241104,Sotiriou:2013qea}.\footnote{This was first shown for stationary black holes in minimally coupled Brans-Dicke theories \cite{hawking1972black}, and subsequently extended to a more general class of scalar-tensor theories including self-interactions of the scalar \cite{PhysRevLett.108.081103}, to spherically symmetric static black holes in Galilean-invariant theories \cite{PhysRevLett.110.241104}, and for slowly rotating black holes in more general shift-symmetric theories \cite{Sotiriou:2013qea}.} Stealth black holes with a constant scalar field background profile $\bar\phi$  have therefore been investigated in \cite{Tattersall:QNMsHorn,Tattersall:BHspectro,Motohashi:2018wdq}.
Around this background odd metric perturbations trivially behave just as in GR, since they are unaffected by the even sector (where scalar perturbations do induce non-trivial effects) and also do not feel any effects from the scalar background solution (since this is trivial in the present no-hair setup).
Consequently, in order to explore potentially observable effects induced by the scalar, one ought to either investigate different background solutions (and hence consider Horndeski theories that evade no-hair theorems) or consider even perturbations. For detailed discussions of the second option we refer to \cite{Kobayashi:2014wsa, Tattersall:QNMsHorn,Tattersall:BHspectro,Tattersall:GenBHPert,Franciolini:2018uyq,Datta:2019npq,Khoury:2020aya,Bernardo:2020ehy,Langlois:2022eta,Minamitsuji:2022mlv,Hui:2021cpm} for work in the context of Horndeski gravity, and to
\cite{Tattersall:GenBHPert,deRham:2019gha,Langlois:2021aji,Glampedakis:2019dqh,Chen:2021pxd,Chen:2021cts,Cano:2023jbk}
for work in the context of other theories (scalar-tensor or otherwise).
However, here we will proceed along the first route, considering the dynamics of odd perturbations around background solutions which evade no-hair theorems.  There exist a number of loopholes around no-hair theorems and one can broadly classify the constructed hairy solutions depending on whether the background scalar is static or contains a (typically linear) time-dependence.
We summarise in Appendix \ref{app-hair} the existence of static radial hair solutions, but focus here exclusively on subsets of Horndeski which admit stealth black hole metrics with a time-dependent scalar as a background solution as in \eqref{backg} of Schwarzschild or Schwarzschild-de Sitter (SdS) form
\begin{align}
    B&=1-\frac{2M}{r}, & &\text{Schwarzschild},\nn\\
    B&=1-\frac{2M}{r}-\frac{1}{3}\Lambda r^2, & &\text{SdS.}
\label{eqn-metric}
\end{align}
Here $M$ is the mass of a spherically symmetric compact object (e.g. a black hole) and $\Lambda$ is a cosmological constant. Note that since we are focusing on stealth solutions where $B(r)$ is given by a GR metric, the scalar hair introduced by $q$ is of secondary nature.\footnote{Non-stealth solutions with primary hair have also been found in \cite{Bakopoulos:2023fmv}, but we will not be considering here.}

 \begin{table*}
  \begin{tabular*}{\textwidth}{@{\extracolsep{\fill}} ccccc}
    \toprule
    \multirow{2}{*}{$\mathcal{L}$} &
      \multicolumn{2}{c}{Background solution} &
      \multicolumn{2}{c}{Stability} \\
      & {$g_{\mn}$} & {$X$} & {odd} & {even} \\
      \midrule
    (Shift + refl)-sym Horndeski \cite{Babichev:2013cya,Kobayashi:2014eva} & $\text{S(dS)}^{*}$ & $\frac{q^2}{2}=\text{const}$  & \cmark & \xmark\\
    Cubic Galileon \cite{Babichev:2012re,Babichev:2016fbg} & $\text{S(dS)}^{*}$ (non-exact) & non-const (non-exact) & \textbf{?} & \textbf{?}\\
    Shift-sym breaking Horndeski \cite{Minamitsuji:2018vuw} & $\nexists$ (for large subclasses) & $\nexists$ (for large subclasses) & \textbf{-} & \textbf{-}\\
    $G_2=\eta X$, $G_4=\zeta+\beta\sqrt{X}$ \cite{Babichev:2017guv} & $\text{S(dS)}^{*}+\text{RN(dS)}^{*}$ (non-exact) & non-const & \textbf{?} & \textbf{?}\\
    Shift-sym beyond Horndeski \cite{Babichev:2016kdt} & $\text{SdS}^{*}$ & $\frac{q^2}{2}=\text{const}$ & \cmark & \xmark\\
    Shift-sym breaking quadratic DHOST \cite{BenAchour:2018dap} & $\text{S(dS)}^{*}$ & $\frac{q^2}{2}=\text{const}$ & \cmark & \xmark\\
    Shift-sym quadratic DHOST \cite{Motohashi:2019sen,Charmousis:2019vnf} & $\text{S(dS)}^{*}+\text{K}^*$ & $\text{const}$ & \cmark & \xmark\\
    Quadratic DHOST \cite{Takahashi:2020hso} & $\text{S(dS)}^{*}+\text{(K)RN(dS)}^{*}$ & $\text{const}$ & \cmark & \xmark\\
    $G_2=-2\Lambda+2\eta\sqrt{X}$, $G_4=1+\lambda\sqrt{X}$ \cite{Bakopoulos:2023fmv} & S(dS) & $\frac{1}{2}\frac{q^2\lambda}{\lambda+\eta r^2}$ & \cmark (this work) & $\textbf{?}$\\
    \bottomrule
  \end{tabular*}
  \caption{Stealth black hole solutions with a linearly time-dependent scalar \eqref{backg}. S(dS) corresponds to Schwarzschild(-de Sitter), while (K)RN(dS) refers to (Kerr-)Reissner-Nordsröm(-de Sitter). The cosmological constant in dS can either come from a bare cosmological constant in the action as in GR, or from an effective combination of beyond-GR parameters (i.e. self-tuned). We denote cases which can fall in the latter category with $*$. The symbol $\nexists$ is used to indicate the non-existence of stealth black hole solutions -- in particular such solutions were shown to be absent in large classes of shift symmetry breaking Horndeski theories in  \cite{Minamitsuji:2018vuw}. Further details related to all the solutions shown in this table are discussed in Appendix \ref{app-hair}. Let us only point out here that even modes have been shown to suffer from instabilities in all known hairy solutions for which $X=\text{const}$
  \cite{deRham:2019gha}. This table builds on a pre-existing one \cite{Motohashi:2019sen}.}
     \label{lit-rev}
\end{table*}
 
Solutions of the form of \eqref{backg} have been found to be of special cosmological interest, where the time-dependent scalar can affect cosmological dynamics and e.g. play the role of dark energy. More generally speaking, when embedding black hole solutions in a cosmological spacetime time-diffeomorphisms are naturally broken in the long-distance limit (unlike for Schwarzschild solutions with Minkowski asymptotics) and hence a a time-dependent solution for the scalar is a natural occurrence in such settings.
Calculating and measuring physical effects arising from such a time-dependence on the gravitational waves emitted by black holes therefore promises to provide informative constraints on models such as \eqref{eq-theory}.
In this work we will focus on the imprints left on the quasinormal modes in the ringdown signal of binary black hole mergers by the non-trivial nature of the scalar field. However, a key requirement that needs to be satisfied prior to carrying out a ringdown study is for the solution to be stable, i.e. to avoid an unphysical (exponential) growth of perturbations.
And indeed, while SdS solutions (i.e. approximants to "cosmological black hole" solutions) have been found for several sections of the Horndeski family, they have also been generically found to suffer from instabilities when the scalar has a time-dependent profile. Table \ref{lit-rev} summarises the existence and stability of these types of solutions, and we provide in Appendix \ref{app-hair} a more in-depth and historical examination of the results collected there. Here, it suffices to say that, while such exact SdS and scalar profile solutions have been found for large sub-classes of Horndeski theories, higher-$\ell$ even modes around backgrounds of the form of \eqref{backg} have been shown to generically suffer from instability or strong coupling issues \cite{deRham:2019gha}.

This then leaves us with no known well-behaved stealth black hole solutions with a linearly time-dependent scalar.
However, as pointed out above, $X=\textit{const}$ is a key requirement for the results of \cite{deRham:2019gha} to hold, so an obvious question is whether solutions with different scalar profiles exist and, if so, whether perturbations can be stable on such other backgrounds where $X\neq\text{const}$.
\footnote{Note that an alternative route to bypass stability issues in specific theories and to study the evolution of perturbations in a model-independent way (detached from the question of whether a given perturbative solution can be embedded in a full covariant theory) is to employ an Effective Field Theory (EFT) formalism to perturbations on a background with a time-dependent scalar. Such formalism has been developed recently in \cite{Mukohyama:2022enj} and employed in different scenarios \cite{Khoury:2022zor,Mukohyama:2022skk,Mukohyama:2023xyf,Konoplya:2023ppx,Barura:2024uog}. In this work, however, we will focus on model-specific exact black hole solutions. In the context of realising well-behaved perturbative setups within covariant theories, also see work on the `scordatura mechanism' \cite{Motohashi:2019ymr,DeFelice:2022xvq}. }
While previous solutions have frequently been constructed by first imposing $X=\textit{const}$ for simplicity (see e.g. a related discussion in \cite{BenAchour:2018dap}) and then finding the form of $\psi$ satisfying this condition, the theory \eqref{eq-theory}  was recently identified as possessing a stealth SdS solution for which $X\neq\textit{const}$ \cite{Bakopoulos:2023fmv}. In terms of Horndeski functions, this theory is given by  
\begin{align}
    G_2&=-2\Lambda+2\eta\sqrt{X}, & G_4&=1+\lambda\sqrt{X}, &G_3 &= 0 = G_5.
    \label{root-X-geom}
\end{align}
As pointed out above, unlike previous solutions, $X$ then adopts a non-trivial radial profile, cf. equations \eqref{backg} and \eqref{eq-psiprofile}. In the next section we will therefore examine this theory at the background and perturbative level, showing that the solution can be stable under odd parity perturbations, and showing the effects on its (likewise odd parity) quasinormal mode spectrum.

\section{Background stability and perturbations}
\label{sec-bh-stab-perts}
Re-expressing \eqref{root-X-geom} as a full action for the theory, we have
\begin{align}
    S= \int d^4x \sqrt{-g}\Big[&R\left(1+\lambda\sqrt{X}\right)-2\Lambda+2\eta\sqrt{X}\nn\\
    &\; +\frac{\lambda}{2\sqrt{X}}\left((\square\phi)^2-(\phi_{\mu\nu})^2\right)\Big],
    \label{root-X-action}
\end{align}
where the second line comes from the $G_{4X}$ term. It is worth making explicit that, just as for the metric determinant, the square root of $X$ is taken to be the principal (positive) square root. Upon a redefinition of the scalar field one can absorb into the scalar field either of the two parameters $\eta$ or $\lambda$ and hence redefine the theory without loss of generality in terms of only one parameter. 
This will be important when investigating and constraining physical deviations from GR in sections \ref{sec-qnms} and \ref{sec:forecasts}. In particular, we will redefine the scalar field as $\phi\rightarrow\frac{\phi}{\eta}$ and use $\beta^2\equiv\frac{\lambda}{2\eta}$ as the single parameter controlling (small) departures from GR. In this section, in order to facilitate comparison with \cite{Bakopoulos:2023fmv}, we will however keep both $\eta$ and $\lambda$ as bookkeeping parameters. We collect in Appendix \ref{app-EOMs} expressions for the covariant equations of motion, the current associated with the theory's shift-symmetry (demonstrating its regularity), and show a relation between the scalar and metric equations.

\subsection{Cosmological background}
We will focus on black hole solutions and related perturbations in this paper and hence primarily investigate \eqref{root-X-action} as a fiducial effective description of physics on the corresponding scales. Nevertheless, given the Schwarzschild-de Sitter stealth solution for the metric, it is interesting to briefly discuss the long-distance, `cosmological' de Sitter limit of the dynamics encoded in \eqref{root-X-action}. 
Focusing on the background evolution in the cosmological $r_s/r \ll 1$ limit, one can map the de Sitter metric in spherically symmetric coordinates -- the long distance limit of \eqref{eqn-metric} -- to standard cosmological coordinates by using the transformations \cite{Babichev:2012re}
\begin{align}
    t&=\tau-\frac{1}{2H}\ln{\left[1-\left(He^{H\tau}\rho\right)^2\right]}\nn \\
    r&=e^{H\tau}\rho,
\end{align}
where $\Lambda = 3 H^2$. This transformation then yields the metric 
\begin{equation}
    ds^2=-d\tau^2+e^{2H\tau}(d\rho^2+\rho^2 d\Omega^2),
    \label{cosmo-coord}
\end{equation}
up to corrections ${\cal O}(r_s/r)$, i.e. terms strongly suppressed in the cosmological long distance limit.
The Hubble parameter $H$
is a constant in the de Sitter limit we are considering here.
In these new coordinates, the canonical scalar kinetic term $X$ is given by
\begin{align}
    X&=\frac{1}{2}\frac{q^2\lambda}{\lambda+e^{2\sqrt{\frac{\Lambda}{3}}\tau}\eta\rho^2} =\frac{e^{-2\sqrt{\frac{\Lambda}{3}}\tau}q^2\lambda}{2\eta\rho^2}+\mathcal{O}\left(\frac{1}{\rho^4}\right),
    \label{eq:Xinrho}
\end{align}
where we again see that $X$ is asymptotically suppressed at large distances.
The fact that the effective cosmological constant is $\Lambda$ (i.e. is not redressed by contributions from the scalar) as well as \eqref{eq:Xinrho} serve to highlight two key points: 1) In the solution we are considering, the scalar field $\phi$ does not affect the cosmological background solution and e.g. is therefore not playing the role of dark energy. While, as discussed above, time-dependent scalar hair is often motivated by the way in which this time-dependence can be linked to cosmological background dynamics, this is therefore not a primary motivation for the specific solution we investigate here. 2) The form of $X$ \eqref{eq:Xinrho} and the fact that it is asymptotically suppressed at large distances means that this is not a solution where one obtains {\it both} $\phi \propto \tau$ in cosmological coordinates {\it and} a metric solution of the simple cosmological form \eqref{cosmo-coord}. Instead, a significant (unsuppressed) $\rho$-dependence remains present in the $\phi$ profile.

Having briefly considered the background evolution, especially in the context of perturbative dynamics it is interesting to note that the theory we are investigating \eqref{root-X-geom} belongs to the class of `extended cuscuton' theories \cite{Iyonaga:2018vnu} -- also see \cite{Afshordi:2006ad,Afshordi:2007yx} for the original cuscuton theory. These theories satisfy the following condition
\begin{equation}
    \mathcal{G}_T\mathcal{K}+6\mathcal{M}^2=0,
    \label{cuscuton-conditon}
\end{equation}
where $\mathcal{G}_T,\mathcal{K}, \mathcal{M}$ are defined for general Horndeski theories in \cite{Iyonaga:2018vnu}, but specialised to our setup are given by
\begin{align}
    &\mathcal{G}_T=2(G_4-2XG_{4X}), \quad\quad \mathcal{M}=2H\partial_\tau\phi(G_{4X}+2XG_{4XX}), \nn \\
    &\mathcal{K}=G_{2X}+2XG_{2XX}+6H^2(G_{4X}+8XG_{4XX}+4X^2G_{4XXX}).
\end{align}
It is straightforward to check that \eqref{root-X-geom} satisfies $\mathcal{K}=0$ and $\mathcal{M}=0$, hence trivially meeting the condition \eqref{cuscuton-conditon}.
For such theories, \cite{Iyonaga:2018vnu} show that the scalar manifestly does not propagate when expressed in coordinates such that $\phi \propto \tau$. While we do not work in a coordinate system where $\phi \propto \tau$ here, the timelike nature of the derivative of $\phi$ means a transformation to new coordinates $\tilde x^\mu$ such that $\phi \propto \tilde\tau$ is expected to exist. Note that the metric is likewise expected to take on a form different from \eqref{cosmo-coord} in this new coordinate system. 
While we will not investigate the cosmological limit in different coordinate systems or further detail here, it will be interesting to investigate the link to cosmological cuscuton analyses further in the future. In the same vein it will also be interesting to in the future go beyond the odd parity perturbations around black hole backgrounds we focus on here. Investigating the even parity sector of \eqref{root-X-geom} will in particular allow an assessment of whether scalar perturbations (which are even parity and hence do not show up in the odd sector) propagate in the black hole solutions considered here and precisely how this connects to the cosmological limit at large distances.\footnote{For a related study of black hole perturbations on a cuscuton-like model see \cite{Saito:2023bhn}, where it is found that in their set up odd parity modes evolve in the same way as in GR. Related to cosmological dynamics, while the model investigated here does not alter the cosmological expansion and dark energy is just given by a cosmological constant, we note that other cuscuton-related models can affect dark energy dynamics \cite{Iyonaga:2020bmm}.} 

\subsection{Black hole background}
\begin{figure}
    \centering
    \centering
    \includegraphics[width = 0.48\textwidth]{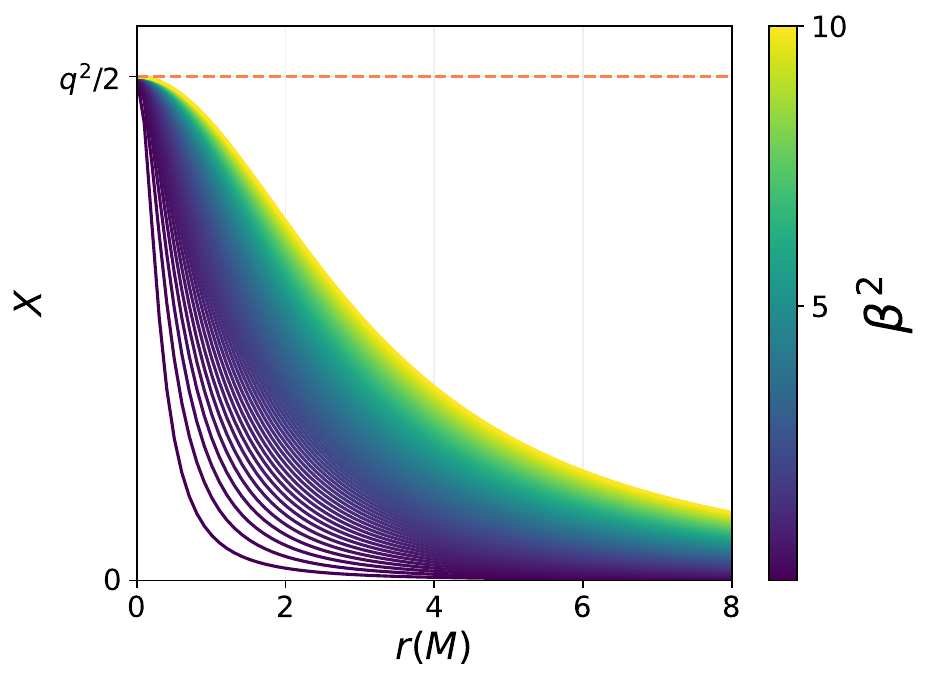}\\[0.5cm]
    \caption{Here we plot $X$ \eqref{X-function}, the standard kinetic term for the background scalar as a function of $r$. We show different choices for $\beta$, defined as $\beta^2=\frac{\lambda}{2\eta}$. Smooth continuous lines exist only for $\beta^2>0$, i.e. $\eta$ and $\lambda$ having the same sign. In fact, it will be shown in section \ref{sec-stabconds} that $\beta^2>0$ is necessary to guarantee stability for all $r$. The horizontal orange dashed line at $X=q^2/2$ corresponds to the limit $\eta=0$.
    }
    \label{fig:X-plot}
\end{figure}

Having briefly discussed cosmological limits, let us now move to the black hole backgrounds which are the main focus of this paper. We recall from equations \eqref{eq-BSdS} and \eqref{eq-psiprofile} that it was shown in \cite{Bakopoulos:2023fmv} that the theory \eqref{eq-theory} has a solution of the form \eqref{backg} with \footnote{We refer to our reproducible Mathematica notebook for the details \cite{ringdown-calculations}.} 
\begin{align}
    &B(r)=1-\frac{2M}{r}-\frac{\Lambda r^2}{3}, \nn \\
    &\psi'^2=\frac{q^2}{B^2}\left(1-\frac{\lambda B}{\lambda+\eta r^2}\right),
    \label{backg2}
\end{align}
where the scalar kinetic term is consequently given by
\begin{align}
    X=\frac{q^2}{2B}-\frac{1}{2}B\psi'^2=\frac{1}{2}\frac{q^2\lambda}{\lambda+\eta r^2}.
    \label{X-function}
\end{align}
As mentioned above, unlike for other known time-dependent solutions, the kinetic term $X$ here is not a constant but rather contains a specific r-dependence (see Figure \ref{fig:X-plot}). Note that the presence of \textit{both} $\eta$ and $\lambda$ is required in order to have $X\neq\textit{const}$.
The form of $\psi$ in \eqref{backg2} ensures that both \eqref{scalarEOMcond} and \eqref{metricEOMcond} are satisfied and thereupon guarantees the existence of stealth black hole solutions such as SdS.

\subsection{Perturbations}
In order to investigate the linear stability of the solution \eqref{backg2} in the theory \eqref{root-X-action} and extract the quasinormal spectrum we need to study the evolution of linear perturbations.
Around the static and spherically symmetric backgrounds considered here such perturbations can be decomposed into odd and even parity perturbations (under rotations), which decouple from one another at linear order (i.e. they evolve independently from one another and can therefore be treated separately).
We will work to leading (linear) order in this paper, but note that this decoupling does not hold at higher orders -- see \cite{Brizuela:2006ne,Brizuela:2007zza,Brizuela:2009qd,Lagos:2022otp,Mitman:2022qdl,Cheung:2022rbm} for details on the behaviour of higher order modes. This linear order decoupling enables us to study the odd sector in isolation. Perturbations on the scalar are purely of even parity and therefore will not be considered here. Odd perturbations are however affected by the background solution they are propagating on, so odd metric perturbations will nevertheless be sensitive to the new physics encoded by the (background solution of the) scalar field $\phi$. 
We split the full metric into background + perturbations as
\begin{align}
    g_{\mu\nu}&=\bar g_{\mu\nu}+h_{\mu\nu},
\end{align}
where $\bar g_{\mu\nu}$ is given by \eqref{backg} and \eqref{backg2}, and $h_{\mu\nu}$ are small perturbations on top of it. 
As we are considering the odd sector and hence no scalar perturbations will be present in our analysis, we will (in an abuse of notation) use the same symbol for the scalar field $\phi$ and its background value.
In the Regge-Wheeler gauge \cite{ReggeWheeler}, these look like
\begin{align}
    h_{\mu\nu}^{\rm odd}=\left(\begin{array}{cccc}
        0 & 0 & 0 & h_0\\
        0 & 0 & 0 & h_1\\
        0 & 0 & 0 & 0\\            
        h_0 & h_1 & 0 & 0
    \end{array} \right)\sin{\theta}\partial_\theta Y_{\ell m},
    \label{hodd}
\end{align}  

where we have set $m=0$ without loss of generality as a consequence of the background metric being static. $h_0$ and $h_1$ are functions of $(r,t)$, where the $t$-dependence will be taken to be of the form $e^{-i\omega t}$.

To obtain the evolution of such linear perturbations, we need to work at quadratic order in the perturbed action, which we define as
\begin{equation}
    S^{(2)}=\frac{1}{2}\int d^4x\sqrt{-g}
\Big[\mathcal{L}_{GR}^{(2)}+\mathcal{L}_{\eta}^{(2)}+\mathcal{L}_{\lambda}^{(2)}\Big],
\label{S2}
\end{equation}
where the expressions for the corresponding quadratic Lagrangians are\footnote{
Note that here we use the standard definition of symmetric and antisymmetric tensors
\begin{align}
    A_{\{\mn\}}&\equiv\frac{1}{2}(A_\mn+A_{\nu\mu}),
    \quad\quad A_{[\mn]}\equiv\frac{1}{2}(A_\mn-A_{\nu\mu}).
    \label{sym-tensors}
\end{align}}
\begin{widetext}
\begin{align}
    \mathcal{L}_{GR}^{(2)}&=\frac{1}{2}\left(\nabla^\sigma h^{\mu\nu}(2\nabla_\nu h_{\mu\sigma}-\nabla_\sigma h_{\mu\nu})+2\Lambda h_{\mu\nu}h^{\mu\nu}\right),\label{L2GR}\\
    \mathcal{L}_{\eta}^{(2)}&=\frac{-\eta}{\sqrt{X}}\left(X h_{\mu\nu}h^{\mu\nu}+\phi^\mu\phi^\nu h_\mu^\sigma h_{\nu\sigma}\right),\label{L2eta}\\
    \mathcal{L}_{\lambda}^{(2)}&=\frac{-\lambda}{2\sqrt{X}}\Big[\nabla_\sigma h_{\mu\nu}\left(X\left(\nabla^\sigma h^{\mu\nu}-2\nabla^\nu h^{\mu\sigma}\right)+\phi^\sigma\phi^\rho\left(\frac{1}{2}\nabla_\rho h^\mn-2\nabla^\nu h_\rho^\mu\right)+\phi^\mu\phi^\nu\nabla_\rho h^{\sigma\rho}+2\phi^\mu\phi^\rho\nabla^{\{\nu}h_\rho^{\sigma\}}\right)\nn\\
    &\hspace{60pt}+h_\mn\bigg(\frac{1}{2}h^\mn((\square\phi)^2-\phi_{\rho\gamma}\phi^{\rho\gamma})+4h_\sigma^\nu(\Lambda\phi^\mu\phi^\sigma-\phi^{\sigma\mu}\square\phi+\phi_\rho^\mu\phi^{\rho\sigma})+4h_{\sigma\rho}\phi^{\mu[\sigma}\phi^{\nu]\rho}\nn\\
    &\hspace{100pt}+\frac{1}{2X}\left(h_\sigma^\nu\phi^\mu\phi^\sigma (\phi_{\rho\gamma}\phi^{\rho\gamma}-(\square\phi)^2)+2\phi^{\sigma\rho}(\phi^\mu\phi^\nu\phi^\lambda_{\{\sigma}h_{\rho\}\lambda}-h_{\sigma\rho}\phi^\mu\phi^\nu\square\phi)
    \right)\nn\\
    &\hspace{100pt}+4\phi^\rho\phi^{\mu\sigma}(2\nabla_{[\sigma}h_{\rho]}^\nu+\nabla^\nu h_{\rho\sigma})+2\phi^\mu\phi^{\sigma\rho}(2\nabla_\sigma h_\rho^\nu-\nabla^\nu h_{\sigma\rho})+2\phi^\mu\phi^{\nu\sigma}\nabla_\rho h_\sigma^\rho\nn\\
    &\hspace{100pt}+2\phi^\sigma\phi_\sigma^\mu\nabla_\rho h^{\nu\rho}+2\phi^\sigma\phi_{\sigma\rho}\nabla^{[\nu}h^{\rho]\mu}+2\square\phi(\phi^\sigma\nabla_\sigma h^{\mu\nu}-2\phi^\mu\nabla_\sigma h^{\nu\sigma}-2\phi^\sigma\nabla^\nu h_\sigma^\mu)\nn\\
    &\hspace{100pt}+\frac{1}{2X}\phi^\mu\phi^\nu\phi^\sigma\left(\phi_\sigma^\rho\nabla_\gamma h_\rho^\gamma+\phi^{\rho\gamma}(2\nabla_{\{\rho}h_{\gamma\}\sigma}-\nabla_\sigma h_{\rho\gamma})\right)
    \bigg)\Big],
    \label{L2lam}
\end{align}
\end{widetext}

Here, we have used the fact that some terms vanish for odd perturbations. e.g. the trace $h\equiv h^\mu_\mu=0$ as can be seen from \eqref{hodd}.\footnote{Note that this is not true for even parity perturbations - we refer to \cite{ringdown-calculations} for full details and recall that we are working in Regge-Wheeler gauge here.} We have used the metric equations of motion for this background, i.e. $R_\mn=\Lambda g_\mn$ and $R=4\Lambda$, and the notation for symmetric and antisymmetric tensors as shown in Appendix \ref{app-EOMs}. From the form of the quadratic action we can already make the following observations: \eqref{L2GR} describes how odd parity modes propagate in GR, \eqref{L2eta} provides modification to the effective potential only while \eqref{L2lam} also provides modifications to the kinetic term.

Substituting the components \eqref{hodd} and the solution for the background scalar \eqref{backg} and \eqref{backg2} into the quadratic action \eqref{S2}, integrating over the angular coordinates and performing several integrations by parts, we can write the action in the following form:
\begin{align}
    S^{(2)}=\int dtdr\Big[&a_1h_0^2+a_2h_1^2+a_3\left(\dot{h}_1^2+h_0'^2-2\dot{h}_1h_0'+\frac{4}{r}\dot{h}_1h_0\right)\nn \\
    &+a_4h_0h_1\Big],
    \label{S2v2}
\end{align}
where a dot and a prime denote derivatives with respect to $t$ and $r$, respectively, and we have dropped an overall multiplicative factor of  $2\pi/(2\ell +1)$ coming from  angular integration. The $a$-coefficients are given by
\begin{align}
    &a_1=\frac{\ell(\ell+1)}{r^2}\left[\left(r\mathcal{H}\right)'+\frac{(\ell-1)(\ell+2)\mathcal{F}}{2B}\right],\nn\\
    &a_2=-\frac{\ell(\ell+1)(\ell-1)(\ell+2)}{2}\frac{B}{r^2}\mathcal{G},\nn\\
    &a_3=\frac{\ell(\ell+1)}{2}\mathcal{H},\nn\\
    &a_4=\frac{\ell(\ell+1)(\ell-1)(\ell+2)}{r^2}\mathcal{J}.
\end{align}
where the $a_i$ are to be evaluated on the background. Expressions for $\{\mathcal{F},\mathcal{G},\mathcal{H},\mathcal{J}\}$ are given by
\begin{align}
    \mathcal{F}&=2\left(G_4-\frac{q^2}{B}G_{4X}\right),\nn\\
    \mathcal{G}&=2\left(G_4+\left(\frac{q^2}{B}-2X\right)G_{4X}\right),\nn\\
    \mathcal{H}&=2(G_4-2XG_{4X}),\nn\\
    \mathcal{J}&=2q\psi'G_{4X},
    \label{FGH}
\end{align}
where in our case the $G$-functions take the form given by \eqref{root-X-geom} but the formula applies to more general theories \cite{Takahashi:2019oxz,Ogawa:2015pea}. Note that $\mathcal{J}\neq0$ only due to the presence of the scalar hair $q\neq0$. The quadratic action (\ref{S2v2}) contains two fields $(h_0,h_1)$, but describes only one dynamical degree of freedom. As shown in \cite{Ogawa:2015pea}, the action can be rewritten to make this manifest.
\begin{align}
    S^{(2)}=\frac{\ell(\ell+1)}{4(\ell-1)(\ell+2)}\int dtdr\Big[&b_1\dot{Q}^2-b_2Q'^2+b_3\dot{Q}Q'\nn \\
    &-(\ell(\ell+1)b_4+V_{\text{eff}}(r))Q^2\Big],
    \label{S2Q}
\end{align}
where the new variable $Q$ can be written in terms of the old ones as
\begin{equation}
    Q=\dot{h}_1-h_0'+\frac{2}{r}h_0
\end{equation}
and the $b$-coefficients are given by
\begin{align}
    b_1&=\frac{\mathcal{F}\mathcal{H}^2r^2}{B\mathcal{F}\mathcal{G}+B\mathcal{J}^2},\nn\\
    b_2&=\frac{B\mathcal{G}\mathcal{H}^2r^2}{\mathcal{F}\mathcal{G}+\mathcal{J}^2},\nn\\
    b_3&=\frac{2\mathcal{H}^2\mathcal{J}r^2}{\mathcal{F}\mathcal{G}+\mathcal{J}^2},\nn\\
    b_4&=\mathcal{H}.
\end{align}

Note the presence of the cross term $b_3$ which includes one time and one radial derivative. By performing a time redefinition as described in \cite{Takahashi:2019oxz,Nakashi:2023vul}
\begin{equation}
    \tilde t=t+\int \frac{b_3}{2b_2}dr,
    \label{t-transf}
\end{equation}
we can diagonalise the kinetic part of the Lagrangian and rewrite the quadratic action in the standard form
\begin{align}
    S^{(2)}=\frac{\ell(\ell+1)}{4(\ell-1)(\ell+2)}\int dtdr\Big[&\tilde{b_1}(\partial_{\tilde{t}}Q)^2-b_2Q'^2\nn \\
    &-(\ell(\ell+1)b_4+V_{\text{eff}}(r))Q^2\Big],
\end{align}
with
\begin{equation}
    \tilde{b_1}=b_1-\frac{b_3^2}{4b_2}=\frac{\mathcal{H}^2r^2}{B\mathcal{G}}.
\end{equation}
The potential is given by
\begin{align}
    V_{\text{eff}}(r)&=r^2\mathcal{H}\left(b_2\left(\frac{1}{r^2\mathcal{H}}\right)'\right)'-2\mathcal{H}.\\
    &=-2\left(r^2\left(\frac{b_2}{r^3}\right)'+\mathcal{H}\right).
    \label{Veff}
\end{align}
where the second equation is obtained using the fact that $\mathcal{H}=2$ in this theory and thus $\mathcal{H}'=0$.

\subsection{Stability conditions}
\label{sec-stabconds}
Having written the quadratic action in the form of \eqref{S2Q}, we can easily identify the following conditions in order for perturbations to be well-behaved and not grow over the background
\begin{align}
    &\tilde{b_1}>0, & &b_2\geq0, & &b_4\geq0.
    \label{b-inequalities}
\end{align}
Since $b_4=\mathcal{H}=2$, the last inequality, which ensures the avoidance of tachyonic instabilities, is always satisfied. Before analysing the first two conditions, it is important to show that they indeed are legitimate measures of stability. As was shown in \cite{Babichev:2018uiw}, the Hamiltonian density in the $(t,r)$ coordinates, i.e. as obtained from \eqref{S2Q}, might be unbounded from below. However, Hamiltonian densities are coordinate-dependent quantities, and by performing a time coordinate transformation of the type \eqref{t-transf} and showing the boundedness from below of the Hamiltonian in the new $(\tilde{t},r)$ coordinates, which is also encapsulated by conditions \eqref{b-inequalities}, it suffices to guarantee stability \cite{Babichev:2018uiw}. In our background of interest, $\tilde{b_1}$ and $b_2$ take the form
\begin{align}
    \label{bs}
    &\tilde{b_1}=2r^2\frac{1}{\frac{|q|}{\sqrt{2}}\sqrt{\lambda^2+\eta\lambda r^2}+B(r)}>0,\nn\\
    &b_2=2r^2\frac{\frac{|q|}{\sqrt{2}}\sqrt{\lambda^2+\eta\lambda r^2}+B(r)}{1+\frac{|q|}{\sqrt{2}}\frac{\lambda^2}{\sqrt{\lambda^2+\eta\lambda r^2}}}\geq0.
\end{align}

Note that $B(r)$ is a positive definite function of $r$, so to satisfy both inequalities and ensure the coefficients are real we require
\begin{align}
    \lambda^2+\eta\lambda r^2\geq0.
    \label{stab-cond}
\end{align}
Investigating this condition in the short and long distance limits and for a fiducial mass of $M=1/2$, one finds
\begin{align}
\begin{cases}
    \lambda^2+\eta\lambda\geq0 & \text{at $r=2M$},\\
    \eta\lambda\geq0 & \text{at $r\rightarrow\infty$}.
    \label{stabconds}
\end{cases}
\end{align}
Note that the second condition is stronger than the first, meaning that $\eta\lambda \geq 0$ is a sufficient condition to guarantee stability for all $r$. It is instructive to point out that we have glossed over the difference between Schwarzschild and SdS metric backgrounds in deriving \eqref{stabconds}. In SdS it does not make sense to take $r \to \infty$, but instead there the cosmological horizon $r_c$ serves as an appropriate long distance limit for $r$. However, $r_c \gg r_s$ by $\sim 20$ orders of magnitude, so corrections to \eqref{stabconds} arising from this finite long-distance limit are very strongly suppressed and we can therefore work with \eqref{stabconds} to very high accuracy even for SdS.    

\begin{figure}
    \centering
    \includegraphics[width = 0.48\textwidth]{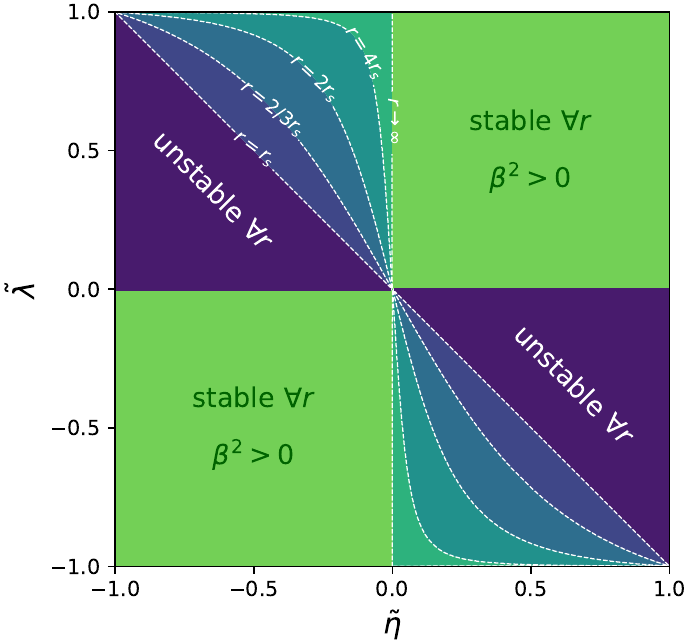}\\[0.5cm]
    \caption{Stability plot for black holes with a time-dependent scalar \eqref{backg} in cuscuton-like theory \eqref{root-X-action}. In the bright green region, stability is ensured for all $r$. At $r=r_s\equiv2M$, the stable region is extended to cover up to the diagonal lines \eqref{stabconds}. In between the two extremes, there is a smooth $r$-dependent transition as shown by the dashed contours. Finally, the dark blue region corresponds to parameter values violating the stability conditions for all $r$. The plot we show here is a 2D representation of what really is a 3D space, where the extra dimension is given by $r$. In \cite{ringdown-calculations} we include an interactive 3D version of this plot which might help in its understanding.}
    \label{fig:stability}
\end{figure}

To visualise the stability conditions in the full parameter space, we can compactify the infinite range of $r$ as well as $\eta$ and $\lambda$, i.e. $-\infty\leq (\gamma,\kappa)\leq\infty$ into a finite range. This can be done with the choices\footnote{Note that inverting this one gets
\begin{align}
    &\eta\equiv\frac{\tilde{\eta}}{\sqrt{1-\tilde{\eta}^2}}, & &\lambda\equiv\frac{\tilde{\lambda}}{\sqrt{1-\tilde{\lambda}^2}}.
\end{align}
}
\begin{align}
    &\tilde{\eta}=\frac{\eta}{\sqrt{1+\eta^2}}, & &\tilde{\lambda}=\frac{\lambda}{\sqrt{1+\lambda^2}}.
\end{align}
Now, the full range of $\tilde{\eta}$ and $\tilde{\kappa}$ is given by $-1\leq (\tilde{\eta},\tilde{\lambda})\leq1$. Figure \ref{fig:stability} summarises the results of our stability analysis using these variables.

As can be easily appreciated from the figure, stability conditions share certain symmetries in the $\eta$ and $\lambda$ plane. This is indeed not surprising, since (as dscussed at the start of this section) one can rewrite the theory we consider in terms of only one free parameter, namely the ratio between $\lambda$ and $\eta$. More specifically, we can redefine the scalar field $\phi\rightarrow \phi/\eta$, which maps the action \eqref{root-X-action} to the following
\begin{align}
    S= \int d^4x \sqrt{-g}\Big[&\mpl^2 R -2\Lambda + 2\mpl^2\sqrt{X} 
    + 2\beta^2 R \sqrt{X} \nn\\
    &\; +\frac{\beta^2}{\sqrt{X}}\left((\square\phi)^2-(\phi_{\mu\nu})^2\right)\Big],
\end{align}
where we have temporarily suspended geometric units to make powers of $\mpl$ explicit and have implicitly defined the parameter $\beta^2$ as
\begin{equation}
    \beta^2\equiv\frac{\lambda}{2\eta} \geq 0,
\end{equation}
where the final inequality is mandated by the stability conditions \eqref{stabconds}. Note that defining the parameter as $\beta^2$ is precisely motivated by those stability conditions, but the square root structure of the theory will mean several predictions in the following sections are controlled by $\beta \equiv \sqrt{\beta^2}$.

\section{Quasinormal modes}
\label{sec-qnms}
Having set up the relevant perturbation theory and discussed stability properties in the previous section, we are now in a position to derive the corresponding quasinormal mode frequencies in the odd sector. As we will see, deviations from standard GR predictions will (as one may expect) be controlled by the $\beta$ parameter introduced above.

\subsection{Modified Regge-Wheeler equation}
In order to obtain the analogue of the Regge-Wheeler equation we will follow the procedures described in \cite{Nakashi:2023vul}. We start by obtaining the equation of motion for the master variable $Q$ of the odd parity perturbations by applying the variational principle to \eqref{S2Q}
\begin{equation}
    -\partial_{\tilde{t}}^2Q+\frac{b_2}{\tilde{b_1}}Q''+\frac{b_2'}{\tilde{b_1}}Q'-\frac{\ell(\ell+1)b_4+V_{\text{eff}}}{\tilde{b_1}}Q=0.
\end{equation}

To remove the single radial derivative term, we express the equation above in a generalisation of the tortoise coordinate given by\footnote{Note that for $\lambda=0$ we recover the tortoise coordinate in GR $dr_*=\frac{1}{B}dr$.}
\begin{equation}
    r_*=\int\sqrt{\frac{\tilde{b_1}}{b_2}}dr
\end{equation}
and redefine $Q$ as
\begin{equation}
    \Psi=FQ
\end{equation}
where
\begin{equation}
    F=\left(\tilde{b_1}b_2\right)^{1/4}.
\end{equation}

Doing this, we obtain an expression in the form of the Regge-Wheeler equation
\begin{equation}
    \left(\partial_*^2-\partial_{\tilde{t}}^2-V\right)\Psi=0,
\end{equation}
where we have expressed the tortoise derivative as $\frac{\partial}{\partial r_*}\equiv\partial_*$. The potential is given by
\begin{equation}
    V=\frac{\ell(\ell+1)b_4+V_{\text{eff}}}{\tilde{b_1}}+\frac{1}{F}\partial_*^2F.
\end{equation}
and $V_{\text{eff}}$ is given by \eqref{Veff}. The full analytical expression for $V$ can be written as
\begin{widetext}
\begin{align}
    V=&V_{RW}\left(1+\frac{q\beta\sqrt{2\beta^2+ r^2}}{B}\right)+\frac{q\beta(B+q\beta\sqrt{2\beta^2+ r^2})
}{4r^2(2\beta^2+ r^2)(2q\beta^{3}+\sqrt{2\beta^2+ r^2})^3}\times\nn\\
    &\;\times\Bigg[q^2\beta^6\left(192\beta^4+104\beta^2r^2+3r^4\right)+2(2\beta^2+r^2)\left(24\beta^4+17\beta^2r^2+2r^4\right)+6q\beta^3\sqrt{2\beta^2+r^2}\left(384\beta^4+416\beta^2r^2+117r^4\right)\nn\\
    &\;-\beta^2B\left(2\left(48q^2\beta^6(2\beta^2+r^2)+48\beta^4+56\beta^2r^2+19r^4\right)\right)+\frac{q\beta^3}{\sqrt{2\beta^2+r^2}}\left(384\beta^4+416\beta^2r^2+117r^4\right)\Bigg].
    \label{Vfull}
\end{align}
\end{widetext}
where $V_{\text{RW}}$ is the well-known Regge-Wheeler potential in GR
\begin{equation}
    V_{\text{RW}}=B\left(\frac{\ell(\ell+1)}{r^2}-\frac{6M}{r^3}\right),
    \label{RWpot}
\end{equation}
and, as discussed above, square roots in \eqref{Vfull} denote principal square roots (this is also the origin of odd powers of $\beta$ seen in the potential).

\begin{figure*}
    \centering
    \includegraphics[width = 0.47\textwidth]{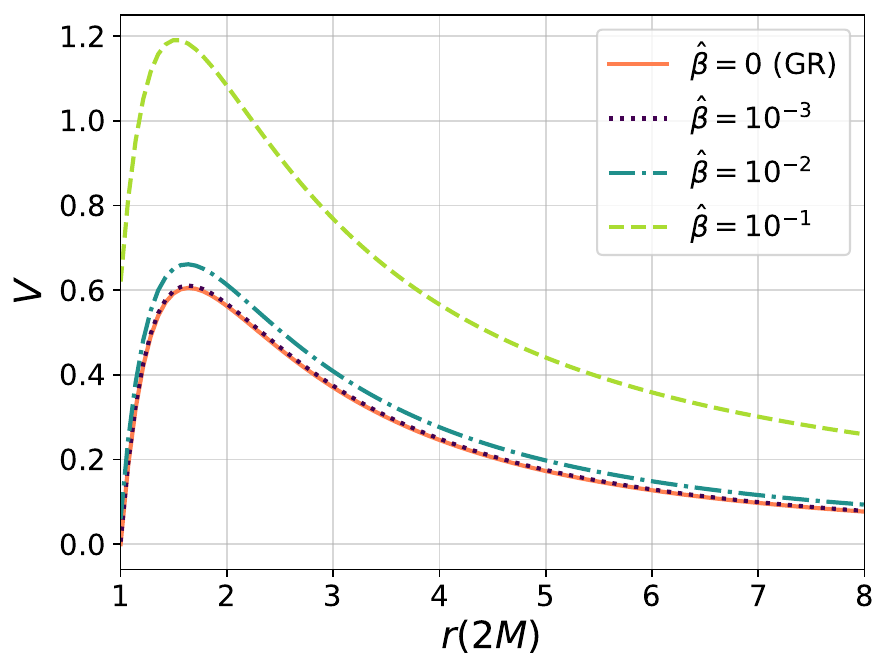}
    \includegraphics[width = 0.49\textwidth]{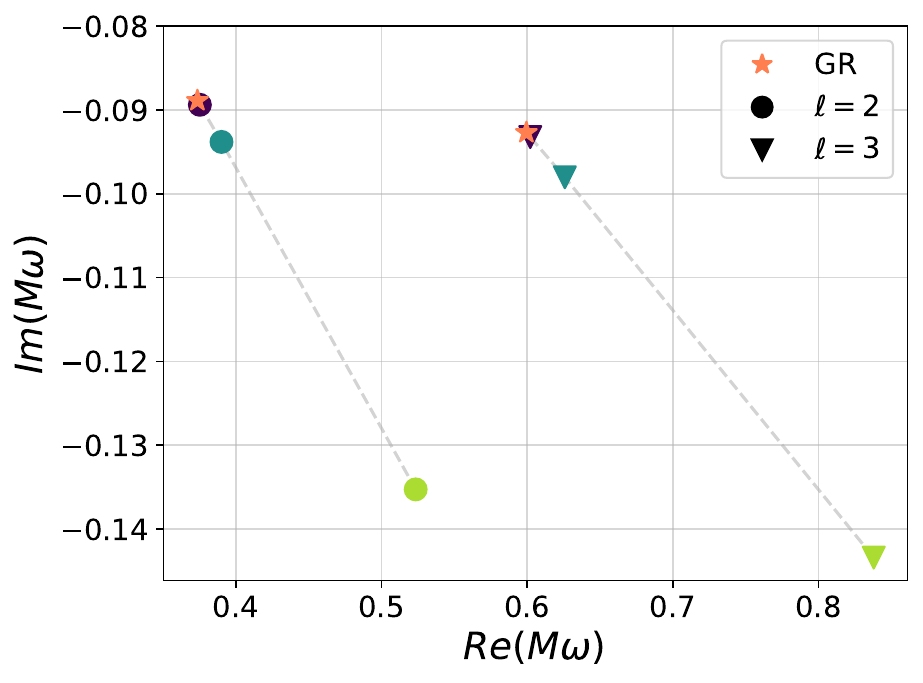}\\[0.5cm]
    \caption{
    {\bf Left panel}: Modified Regge-Wheeler potential for different magnitudes of the deviation parameter $\hat\beta$ for $\ell=2$. We show $r$ in units of $r_s=2M$, so that $r=1$ corresponds to $r=r_s$.
    One can observe that the main effect is an enhancement in the overall amplitude.
    Also note that the maximum of the potential is slightly shifted to lower values of $r$. {\bf Right panel}: Quasinormal modes for $\ell=2$ (corresponding to table \ref{QNM-table}) and $\ell=3$ for increasing magnitudes of $\hat\beta$. Both stars correspond to the GR values and colours correspond to $\hat\beta$ values on the left panel.}
    \label{fig:V-potential-QNMs}
\end{figure*}

In the next section we will investigate the quasinormal frequencies and damping times of this potential. There, we will find that, at leading order, deviations from GR are in fact controlled by a single effective parameter, namely 
\begin{align}
    \hat\beta \equiv \beta \frac{M q}{\Mpl^3}.
\end{align}
Let us now look at how the potential $V$ looks analytically when $\hat\beta \ll 1$, so as to better understand its parametric form. To this end it is instructive to write the potential as
\begin{align}
    V=V_{\text{RW}}+\sum_{i=1}\delta V_i\times\hat\beta^i.
\end{align}
with the first orders of $\delta V_i$ given by
\begin{align}
    &\delta V_1=\frac{1}{M r}\left(\mpl^2(\ell(\ell+1)+1)-\frac{8M}{r}\right),\nn\\
    &\delta V_2=\frac{\mpl^4}{M^2},\nn\\
    &\delta V_3=\frac{\mpl^2}{q^2M^3 r^3}\left(\mpl^4(\ell(\ell+1)-4)+\frac{21M\mpl^2}{r}-\frac{38M^2}{r^2}\right),\nn\\
    \label{potential-coeffs-nu}
    &\delta V_4=\frac{\mpl^6}{q^2M^4 r^2}\left(\frac{56M}{r}-15\mpl^2\right),
\end{align}
where we have again temporarily suspended geometric units to make powers of $\mpl$ (and hence mass dimensions) explicit. Higher orders can be found in the companion Mathematica notebook \cite{ringdown-calculations}.
The effect of $\delta V$ on the unperturbed $V_{RW}$ is shown in Figure \ref{fig:V-potential-QNMs}, where one can appreciate almost a constant shift in the amplitude of the potential throughout the range of the radial coordinate.

\subsection{Quasinormal frequencies}
Having obtained the effective potential for the modified Regge-Wheeler equation, we can now study its quasinormal mode solutions. Looking for quasi-normal mode solutions where $\Psi$ has a time dependence of $e^{-i\omega\tilde{t}}$, we can write the Regge-Wheeler equation as
\begin{equation}
    \left(\partial_*^2+\omega^2-V\right)\Psi=0.
    \label{RWeq}
\end{equation}
Upon imposing the dissipative boundary conditions
\begin{align}
    \Psi\sim\begin{cases}
        e^{-i\omega r_*}, & \text{for} \;\; r_*\rightarrow-\infty,\\
        e^{i\omega r_*}, & \text{for} \;\; r_*\rightarrow\infty,
    \end{cases}
    \label{boun-conds}
\end{align}
which correspond to outgoing waves at spatial infinity and ingoing waves at the black hole horizon, the $\omega$ solutions to \eqref{RWeq} become complex and can therefore be written as
\begin{align}
    \omega_{\ell m}=2\pi f_{\ell m}+\frac{i}{\tau_{\ell m}}.
    \label{omega}
\end{align}
Here, the real part corresponds to the physical oscillation frequency of a mode, while the imaginary part corresponds to its damping time. Due to the axisymmetric nature of our background modes with different $(\ell,m)$ indices do not mix.\footnote{There is a third index characterising the QNM spectrum, the overtone number $n$. Here we only focus on the `fundamental mode' $n=0$. Modes with higher $n$'s (i.e. overtones) are more suppressed by virtue of having increasing values of $|\text{Im}(\omega)|$.} Here we will mainly focus on the dominant mode $(2,2)$.

There are a number of techniques one can use to obtain the QNM themselves (see e.g. \cite{Ferrari:1984zz,1972ApJ...172L..95G,1985ApJ...291L..33S,Motl:2003cd,Motl:2003cd,Horowitz:1999jd,Berti:2009wx,Leaver:1985ax,ParamRingdown}) but we refer to \cite{Berti:2009kk} for an extensive review of those. Here, we will use the WKB method, first applied to black holes in \cite{1985ApJ...291L..33S} and subsequently extended to higher orders in \cite{Iyer:1986np,Konoplya:2003ii,Matyjasek:2017psv,Konoplya:2019hlu}.\footnote{Because of the form of the potential corrections \eqref{potential-coeffs-nu},
in particular the fact that they do not scale with $B(r)$, it is unfortunately challenging to use the parametrised ringdown formalism introduced in \cite{ParamRingdown} and subsequently developed in \cite{McManus:2019ulj,Kimura:2020mrh,Franchini:2022axs}.} We do this by adapting the Mathematica package which can be found in \cite{Konoplya:2019hlu,wkb-package}.

The WKB method provides a straightforward and precise technique of obtaining quasinormal frequencies from an effective potential in a semi-analytical manner. However, in order for it to be justifiably applied, the potential needs to satisfy some criteria laid out in \cite{Konoplya:2019hlu}. Most notably, the potential must have one local maximum, be asymptotically constant, and contain two turning points. As can be observed form Figure \ref{fig:V-potential-QNMs}, these are all satisfied by our modified effective potential.\footnote{Note that, as is also the case for the GR potential in \ref{fig:V-potential-QNMs}, the second turning point for the modified potentials is hidden at $r<1$, where $V$ approaches $+\infty$ as $r\rightarrow0$.}

Let us now briefly introduce how the method works practically. The rationale is to match the asymptotic solutions given by \eqref{boun-conds} respectively with a Taylor expansion around the maximum of the potential located at $r_0$. For a differential equation written in the form of \eqref{RWeq}, the matching of solutions at the different regions imposes
\begin{align}
    \omega^2=V_0-i\left(n+\frac{1}{2}\right)\sqrt{2V_0^{(2)}}-i\sqrt{2V_0^{(2)}}\sum_{i=2}^N\Lambda_j.
    \label{wkb-formula}
\end{align}

Here, $V_0$ denotes the potential evaluated at the maximum $r_0$, and we use $V_0^{(2)}$ to denote the second tortoise derivative of $V$ evaluated at the maximum. $n$, being the overtone number, is set to zero when focusing on the fundamental mode. Finally, $\Lambda_j$ are functions of higher order derivatives of the potential, where $j$ denotes the order to which the WKB expansion is carried out. The first application of this method to black holes was carried out in \cite{1985ApJ...291L..33S} and included only the first order. \cite{Iyer:1986np} then extended the WKB formula to 3rd order by computing $\Lambda_2$ and $\Lambda_3$. This was then extended in \cite{Konoplya:2003ii} to 6th order and in \cite{Matyjasek:2017psv} to 13th order. However, going up in WKB order does not necessarily mean improving accuracy. For instance, the fundamental mode in GR is best approximated by 6th order WKB \cite{wkb-package}, and we find that this is also true when including our corrections.\footnote{We show in \cite{ringdown-calculations} that this is the case by computing how error estimation increases with WKB order. As said, we find that this is minimised for 6th order.} Therefore, we perform calculations to 6th order WKB in what follows.

Note that obtaining QNMs via the WKB method involves taking derivatives at the maximum of the potential. The location of the maximum, however, changes as a function of $\hat\beta$. The WKB package \cite{wkb-package} allows one to obtain numerical values for the QNMs while taking this into account automatically, and we show some examples in table \ref{QNM-table} and figure \ref{fig:V-potential-QNMs} calculated this way.\footnote{We thank Simon Iteanu for comments on the correct implementation of the package.}
 \begin{table}[t!]
 \setlength{\tabcolsep}{11pt}
    \centering
    \sisetup{
        table-number-alignment = left,
        round-mode = places,
        round-precision = 4,
        table-format = +1.3,
        detect-weight = true,
        detect-family = true
    }
    \begin{tabular}{c S[table-format=-1.2] S[table-format=-1.2]}
        \toprule
        & $\text{Re($M\omega$)}$ & $\text{Im($M\omega$)}$\\\midrule
          $\hat\beta=0$ (GR)         & 0.373619 & -0.0888910
          \\
          $\hat\beta=10^{-3}$         & 0.375292 & -0.0893873
          \\
        $\hat\beta=10^{-2}$         & 0.390134  & -0.0938158
        \\
        $\hat\beta=10^{-1}$         & 0.523444  & -0.135257
        \\\bottomrule
    \end{tabular}
    \caption{Real and imaginary components of the $\ell=2$ quasinormal frequencies $\omega$ for varying $\hat\beta$ obtained with the WKB method to 6th order.}
     \label{QNM-table}
\end{table}

\begin{figure}
    \centering
    \includegraphics[width = 0.48\textwidth]{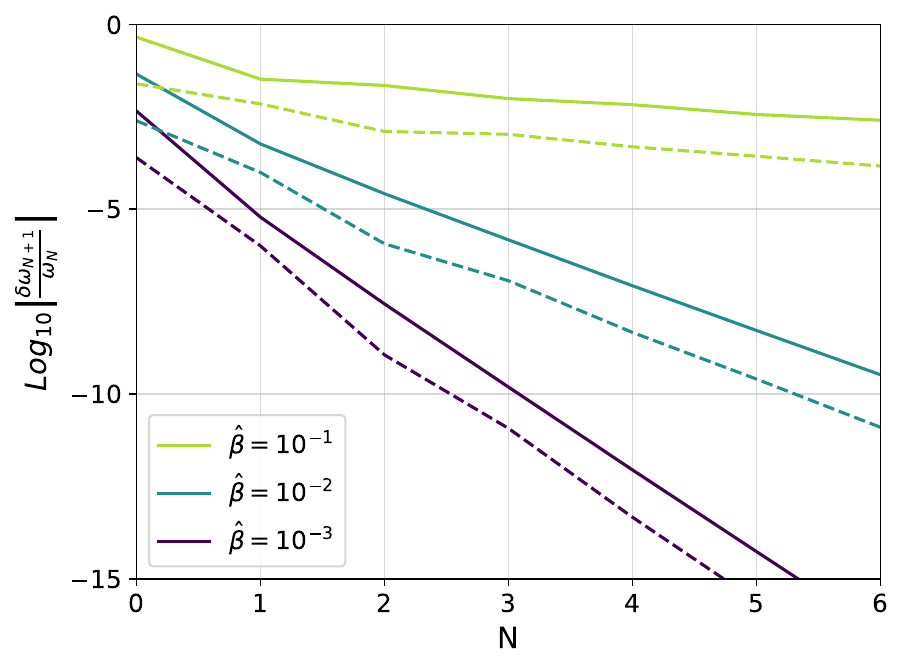}\\[0.5cm]
    \caption{This plot displays the achievable accuracy in the real (solid) and imaginary (dashed) parts of quasinormal modes for $\ell=2$ for increasing orders of $N$ in \eqref{omei}. The y-axis represents the fractional contribution to the quasinormal frequencies gained by going to the next $N+1$ order as compared to $\omega_N$. As can be seen, these fall of quickly, resulting e.g. in \red{${\cal O}(10^{-10})$ corrections for $N \geq 6$ for $\hat\beta \sim 0.01$.}
    }
    \label{fig:error-qnm}
\end{figure}

Nonetheless, we also want `semi-analytical' expressions for $\omega$ as a function of $\hat\beta$ that we can then use in our forecast analysis. For this, we employ the \textit{light ring expansion} \cite{Franciolini:2018uyq} to find the location of the maximum as a function of $\hat\beta$. The expansion works under the assumption that the geometry is "quasi-Schwarzschild" or, in other words, that the location of the potential maximum is only a small deviation away from the Schwarzschild value $r_*^{\text{max}}=r_*^{\text{max, GR}}+\delta r_*$. This is certainly true in our case of study with small $\hat\beta$, as can be seen from figure \ref{fig:V-potential-QNMs}.\footnote{Note that the maximum is approximately located around the light ring, which in Schwarzschild GR is at $r=3M$. In figure \ref{fig:V-potential-QNMs}, because $r_s=1$ has been chosen, the three maxima appear around $r(2M)\approx1.5$.} Upon employing the light ring expansion we can expand the 
defining property of the maximum of the potential $\delta_r V|_{r=r_*^{max, GR}+\delta r_*}=0$ to approximate to first order \cite{Franciolini:2018uyq}
\begin{equation}
    \delta r_*=-\frac{\partial_rV}{\partial_r^2V}\Big|_{r=r_*^{max, GR}}.
\end{equation}
With this, we then have expressions for the maximum as a function of $\hat\beta$ that we can then substitute in all the potential derivatives in \eqref{wkb-formula}. While the full expression of $r_*^{\text{max}}(\hat\beta)$ is quite lengthy -- see \cite{ringdown-calculations} for full details -- this simplifies considerably for small $\hat\beta$, i.e. the case we are focusing on here. But rather than immediately truncating to the lowest order correction in $\hat\beta$, it is instructive to examine the few lowest order terms. We find
\begin{align}
    \frac{\Mpl^2}{M} r_*^{\text{max}}
    &=3.2808 - 3.0306 \cdot \hat\beta +0.8316 \cdot \hat\beta^2 \nn\\
    &-\left(0.22819 + 1.8502 \frac{\Mpl^8}{M^4 q^2}\right)\cdot \hat\beta^3 + {\cal O}(\hat\beta^4)
    \label{eq_rmax}
\end{align}
where we have set $\ell = 2$ in deriving this expression, and the first term represents the GR value $r_*^{\text{max, GR}}=3.2808 \cdot M/\Mpl^2$.
This expression highlights three important points:
\begin{itemize}
    \item[1.] First, the regime where it makes sense to truncate the above expansion is $\hat\beta \ll 1$. This corresponds to $\beta M q \ll \Mpl^3$, which puts an implicit bound on the (until now unrestricted) parameter $q$ if we are to demand working in the small $\hat\beta$ regime, i.e. in the regime where potential deviations and the shift of $r_\star^{\rm max}$ are small. 
    As an example, taking $\beta \sim {\cal O}(1)$ and considering black holes with $M \sim {\cal O}(10) M_\odot$, where $M_\odot \sim 10^{30}\text{kg}$ while $\Mpl \sim 10^{-8}\text{kg}$, this requirement becomes $\sqrt{q} \ll 10^{-28}\text{kg}$ or, equivalently, $\sqrt{q} \ll 10^8 \; \text{eV}$.
    \item[2.] At cubic order in $\hat\beta$ we effectively see that the theory ultimately is controlled by two parameters (in addition to $M$ and $\Mpl$): $\beta$ and $q$. While at lower (leading) orders these only enter in the form of the single effective parameter $\hat\beta$, from cubic order onwards we can see $q$ entering independently. Note that this is expected in light of the potential corrections \eqref{potential-coeffs-nu}, which share this feature. If we require these qualitatively different terms (e.g. the second term in the second line of \eqref{eq_rmax}) to be suppressed with respect to the solely $\hat\beta$-dependent terms, we in addition require $\Mpl^8 \ll M^4 q^2$. Taking the same example masses as above, this is akin to requiring $\sqrt{q} \gg 10^{-47} \; \text{kg}$ or $\sqrt{q} \gg 10^{-11} \; {\rm eV}$, when this bound is satisfied, these additional terms can be safely dropped for black hole mass ranges close to the example chosen here.
    \item[3.] For small $\hat\beta$, the linear $\hat\beta$ term in \eqref{eq_rmax} is the leading order correction. An immediate consequence of this is that the maximum of the potential always decreases in such cases, getting closer to the light ring at $r=3M$ compared to GR. 
\end{itemize}
\black{Equipped with \eqref{eq_rmax}, we can substitute this into the WKB formula \eqref{wkb-formula} to obtain an expression for the QNM frequencies in terms of $\hat\beta$, i.e. $\omega(\hat\beta)$. Since the full expression is quite cumbersome, we do not show it here (but leave it available in \cite{ringdown-calculations} for the interested reader). We have checked that the $\omega$ values found with this analytical expression agree with those in table \ref{QNM-table}.
Working in a small $\hat\beta$ expansion, one then finds
\begin{align}
    \frac{M \omega}{\Mpl^2} = &\frac{M \omega_0}{\Mpl^2} +(1.68-0.50 i) \hat\beta \nn \\
    &-(2.39-0.16 i) \hat\beta^2 + {\cal O}(\hat\beta^3),
\end{align}
where $\omega_0$ is the GR prediction, satisfying $M \omega_0/\Mpl^2 = (0.37-0.09 i)$, and the other terms correspond to entries in \ref{QNM-exp-table}. More precise and extensive coefficients for this expansion are given in this table, with corrections to the QNM frequencies parametrised via
\begin{equation}
    \omega_N=\omega_0+\sum_{i=1}^{i=N}\delta\omega_i\hat\beta^i.
    \label{omei}
\end{equation}
The numerical precision of this expansion is shown in figure \ref{fig:error-qnm} for increasing $N$.}


\begin{table}[ht]
 \setlength{\tabcolsep}{11pt}
    \centering
    \sisetup{
        table-number-alignment = center,
        round-mode = places,
        round-precision = 4,
        table-format = +1.3,
        detect-weight = true,
        detect-family = true
    }
    \begin{tabular}{c S[table-format=-1.4] S[table-format=-1.2] S[table-format=-1.6] S[table-format=-1.2] S[table-format=-1.6]}\toprule
        \multirow{2}{*}{$i$} &
      \multicolumn{2}{c}{$M\delta\omega_i$} &
      \multicolumn{2}{c}{$M\omega_i(\hat\beta=10^{-1})$} \\
      & $\text{Re}$ & $\text{Im}$ & $\text{Re}$ & $\text{Im}$ \\
      \midrule
    0 & / & /  & 0.3736 & -0.0889\\
    1 & 1.67503 & -0.496737 & 0.541122 &-0.138565\\
    2 & -2.38548 & 0.16408 & 0.51727  & -0.13692\\
    3 & 10.40324 & -2.02190 & 0.52767 & -0.13895\\
    4 & -59.1336 & 9.4007 & 0.52176  & -0.13801\\
    5 & 337.0564 & -61.8351 & 0.52513  & -0.13862\\
    6 & -2076.7556  & 394.5525 & 0.52305  & -0.13823\\
    7 & 13129.5241  & -2637.0521 & 0.52436 & -0.13849 \\
        \bottomrule
    \end{tabular}
    \caption{Real and imaginary components of the corrections to the $\ell=2$ quasinormal frequencies $\omega$.}
     \label{QNM-exp-table}
\end{table}

\section{Forecasted constraints} \label{sec:forecasts}
In the previous section we obtained the QNM spectrum for a hairy black hole with a time-dependent scalar, where deviations from GR are controlled by the parameter $\hat\beta$, which encodes information about the underlying scalar-tensor theory.
Our aim now is to forecast how well current and future GW experiments will be able to constrain $\hat\beta$ using solely the ringdown. Table \ref{tab-SNRringdown} collects our main results.

We employ a Fisher forecast analysis to estimate the {precision with which $\hat\beta$ will be measurable}. 
Our analysis ubiquitously uses techniques developed in \cite{Berti_2006} - for a detailed summary of the forecasting setup specialised to our present analysis see \cite{Lahoz:2023csk}.
We will here focus on a simplified scenario where all the standard waveform parameters are already known $(A,\phi^+,...)$, leaving $\hat\beta$ as the only free parameter. {Given this idealised setup, the constraints forecasted here should be interpreted as optimistic/optimal estimates for expected achievable precision. Forecasts with full joint constraints for all waveform parameters will be left for future work.}

For a setup as considered here, where the only waveform parameters we want to constrain are those appearing inside the quasinormal frequencies $\omega$, \black{general expressions for the achievable precision can be derived analytically. These only depend on the number of parameters one wants to constrain.} Here, GR deviations are solely controlled by the parameter $\hat\beta$ so we employ the expression for single-parameter constraints
\begin{equation}
    \sigma_{\hat\beta}^2\rho^2=\frac{1}{2}\left(\frac{f}{Qf'}\right)^2,
    \label{error}
\end{equation}
where the prime denotes a derivative with respect to $\hat\beta$ -- {we refer to \cite{Lahoz:2023csk} for details on the derivation of \eqref{error}}. In \eqref{error}, $\rho$ refers to the standard signal-to-noise-ratio (SNR) and the quality factor $Q$ is defined as
\begin{align}
    Q_{\ell m}=\pi f_{\ell m}\tau_{\ell m}.
    \label{Q-factor}
\end{align}

In the context of astrophysical binary compact objects, the $(\ell,m)=(2,2)$ mode is generically the one with largest amplitude and dominates the ringdown signal \cite{Berti_2006,London:2014cma,Berti:2007fi,Berti:2007zu,Bhagwat:2019bwv,Bhagwat:2019dtm}. It is therefore usually referred to as the dominant mode. For non-rotating black holes, which are our focus here, the equations of motion do not depend on $m$ \cite{Tattersall:GenBHPert}. Consequently, even though $m=0$ is typically set for simplicity, as we have done here, the results are valid for any $m$. For discussions on the amplitude and detectability of subdominant modes we refer to \cite{Berti:2007zu,London:2014cma,Kamaretsos:2011um,Baibhav:2020tma}, but here highlight that not only does the $\ell=2$ mode dominate in typical scenarios, but also note that for binary systems which have orbited each other sufficiently long, the orbits will have circularised, additionally enhancing the $\ell=2$ mode relative to other modes \cite{LIGOScientific:2018jsj,PhysRev.136.B1224,PhysRevD.77.081502}. That being said, our analysis here can straightforwardly be repeated for higher $\ell$ modes,\footnote{Note that the dipole mode $\ell=1$ requires special treatment, as the Regge-Wheeler gauge does not completely fix the gauge in this case. For detailed information see \cite{Kobayashi:2012kh}, where it is also shown that $\ell=1$ contributions are negligible in set ups as the one considered here.} {while we leave conducting a multi-mode analysis 
and extending this study to slowly rotating black holes (see e.g. \cite{Tattersall:KerrdSBH,Pani:2012bp,doi:10.1142/S0217751X13400186,Pani:2013wsa,Brito:2013wya}) for future work.} \red{Note that, in the context of constraints purely derived from ringdown, in order to use black hole spectroscopy to both constrain the standard GR black hole parameters (mass and angular momentum), as well as additional beyond-GR parameters (such as $\hat \beta$ here), the measurement of at least two QNMs is required. This means an additional (subdominant) mode is required in addition to the $(2,2)$ mode discussed above, which typically entails an order of magnitude reduction in the constraining power, see e.g. \cite{Branchesi:2023mws}. So when using a single mode analysis here, this is based on the idealised assumption discussed at the start of this section, that all parameters except for $\hat\beta$ are (already) known to sufficient accuracy (e.g. from the inspiral), and hence only one mode is needed to constrain the single remaining free parameter.}

We can now use the semi-analytical expressions found for the $\hat\beta$-dependent QNMs together with the error expression \eqref{error} to place estimated order-of-magnitude constraints on $\hat\beta$ and hence on the gravity model \eqref{eq-theory}.
Reading off $f$ and $Q$ from \eqref{omei}, as defined in \eqref{omega} and \eqref{Q-factor}, and substituting them into the single-parameter error expression \eqref{error}, we obtain an estimate on its detectability in the same fashion as \cite{Tattersall:QNMHair,Lahoz:2023csk}.\footnote{Note that, in evaluating the final expression, we set $\hat\beta$ to zero. This should simply be understood as capturing the leading order contributions to the error -- depending on the actual value of $\hat\beta$ the precise error can differ by $ \lesssim {\cal O}$(1\%) for $\hat\beta\lesssim 10^{-1}$.}
This gives us\footnote{A value with more significant figures is provided in \cite{ringdown-calculations}. Ultimately, we will approximate $\sigma_{\hat\beta}\rho\approx {\cal O}(10^{-1})$ in table \ref{tab-SNRringdown}, as we are purely interested in the robust order-of-magnitude constraints here.}
\begin{equation}
    \sigma_{\hat\beta}\rho\approx 0.08.
    \label{eq:betaSNR}
\end{equation}
The precision with which $\hat\beta$ can be constrained therefore inversely depends on the achievable SNR, which varies for different existing and upcoming detectors. Table \ref{tab-SNRringdown} collects updated estimates on the optimistic obtainable ringdown SNRs together with the corresponding order-of-magnitude constraint on $\hat\beta$ for several ground and space-based detectors. $3^{rd}$ generation space based detectors such as LISA are predicted to be able to achieve a ringdown SNR as high as \red{$10^3$}, which would entail a constraint on $\hat\beta$ of
\begin{align}
    \sigma_{\hat\beta}^{\rm LISA/TianQin} \sim \red{10^{-4}}.
\end{align}

Let us stress here that the primary bounds discussed in this analysis are projected from a single ringdown observation with the SNR as described in Table \ref{tab-SNRringdown}. In fact, as the number of equivalent events $N$ increases, such bounds are predicted to improve as $N^{1/2}$ \cite{Yang:2017zxs,Brito:2018rfr}. For the LISA band, this could entail an improvement in the constraint on $\hat\beta$ of up to two orders of magnitude, i.e. $\sigma_{\hat\beta}^{\rm LISA/TianQin} \sim \red{10^{-6}}$, as the estimated rates of SMBH mergers, despite somewhat uncertain, lie in the ${\cal{O}}(10-100)$ per year range -- see e.g. \cite{Berti:2006ew,Sesana:2004gf,Rhook:2005pt,Tanaka:2008bv,Berti:2009kk,eLISA:2013xep,Bonetti:2018tpf,Erickcek:2006xc}.\footnote{This however assumes an optimistic scenario of $N$ events with identical SNR, while many events will have lower SNR values, e.g. mergers at higher redshifts.}
\red{Having pointed out how constraints may improve going forward, let us however also re-emphasise that the bounds quoted are from an idealised analysis using only the dominant ringdown mode to constrain $\hat\beta$, with other (GR) parameters inferred to high accuracy from the inspiral and merger phases. As discussed above, a pure ringdown constraint would require the measurement of another subdominant ringdown mode, typically entailing an order of magnitude reduction in the constraining power \cite{Branchesi:2023mws}.}
Finally, let us point out that shifts in the QNM frequencies and damping times are detectable with approximately the same precision as constraints on $\hat\beta$, e.g. $\delta\omega \sim \red{10^{-4}}$ will be detectable with LISA/TianQin. In our context this can e.g. be seen by noticing that the precision with which $\hat\beta$ is constrainable is inversely proportional to SNR \eqref{eq:betaSNR}, while the shift in QNM frequencies and damping times scales linearly with $\hat\beta$ (in the small $\hat\beta$ regime we are investigating) at leading order with order one coefficient(s) $\delta\omega_1 \sim 1.68 - 0.50 i$.

\begin{table}[!t]
\centering
\begin{tblr}{
  colspec = {|c|c|c|}
}
\hline[1pt]
   Detector(s) & Ringdown SNR ($\rho$) & Error on $\hat\beta$\\
   \hline[1pt]\hline[1pt]
    LVK & $10$ \cite{TheLIGOScientific:2016src,Flanagan:1997sx,Nakano:2021bbw}
    & $10^{-2}$\\
   \hline[1pt]
       ET / CE & $10^2$ \cite{Maggiore:2019uih,Nakano:2021bbw,Evans:2021gyd,Hall:2022dik} & $10^{-3}$\\
   \hline[1pt]\hline[1pt]
    pre-DECIGO & $10^2$ \cite{Nakamura:2016hna} & $10^{-3}$\\
   \hline[1pt]
    DECIGO / AEDGE & $10^3$ \cite{Nair:2018bxj,AEDGE:2019nxb}* & $10^{-4}$\\
    \hline[1pt] \hline[1pt]
    LISA & \red{$10^3$} \cite{Bhagwat:2021kwv,Yi:2024elj,PhysRevD.100.044036,Shi:2024ttu} & \red{$10^{-4}$}\\
   \hline[1pt]
    TianQin & \red{$10^3$} \cite{PhysRevD.100.044036,Shi:2024ttu} & \red{$10^{-4}$}\\
   \hline[1pt]
    AMIGO & $10^5$ \cite{Baibhav:2019rsa} & $10^{-6}$\\
   \hline[1pt]
\end{tblr}
\caption[ringdownSNR]{Achievable order-of-magnitude ringdown SNRs for a single observed event for different GW detectors and the corresponding order-of-magnitude errors on $\hat\beta$. A star * denotes that the quoted forecasted SNR is not ringdown-specific. For ET/CE we have quoted the ringdown-specific ET forecast \cite{Nakano:2021bbw}, in the current absence (to our knowledge) of an analogous forecast for CE. \red{For LISA, we quote the maximum SNR estimates from \cite{Bhagwat:2021kwv,Yi:2024elj,PhysRevD.100.044036,Shi:2024ttu} going up to $\sim {\cal O}(10^3)$, which coincide with the SNRs in the LISA Mock Data Challenge \cite{Baghi:2022ucj,MockLISADataChallengeTaskForce:2009wir}, but also note that some more optimistic SNR forecasts (up to $\sim {\cal O}(10^5)$) exist for sufficiently nearby and massive events \cite{Flanagan:1997sx, Zhang:2021kkh}. The same applies for TianQin.} This also illustrates that there is still significant variance in the forecasted SNRs relevant for the missions considered here.}
    \label{tab-SNRringdown}
\end{table}

\section{Conclusions} \label{sec:conclusions}
In this paper we have investigated the stability and quasinormal modes of odd parity perturbations of hairy black hole solutions with a linearly time-dependent scalar. In particular, we have focused on the specific scalar-tensor theory identified in \cite{Bakopoulos:2023fmv} and given by \eqref{eq-theory}. This theory possesses an exact Schwarzschild-de Sitter solution for the metric (\eqref{eq-psiprofile} and \eqref{backg}) and the scalar field is indeed linearly time-dependent for this solution, but (unlike for other currently known exact solutions in scalar-tensor theories with these characteristics) the canonical scalar kinetic term  $X\equiv-\frac{1}{2}\phi_\mu\phi^\mu$ is not constant. This solution is of particular interest since known constant $X$ solutions are generically plagued by instabilities.
Our key findings are:
\begin{itemize}
    \item The solution we investigate can support stable odd parity perturbations, where a non-trivial bound on the theory's coupling constant $\hat\beta^2 > 0$ is placed by requiring odd sector stability. This parameter is effectively a dimensionless measure of the strength of interactions in the theory (and suggestively written as $\hat\beta^2$, given this bound - note, however, that the square root structure of the theory means several theory predictions are controlled by $\hat\beta \equiv \sqrt{\hat\beta^2}$).
    \item We have derived the modified Regge Wheeler equation governing odd parity perturbations in this setup, and provided its quasinormal mode solutions. QNMs in this setting are controlled by the same single parameter $\hat\beta$ highlighted above. We have computed the QNM spectrum and quantified deviations from the GR values as a function of $\hat\beta$, as illustrated in Figure~\ref{fig:V-potential-QNMs}, finding the following qualitative features: 1) The maximum of the modified Regge-Wheeler potential shifts to smaller radii, lying closer to the light ring than in GR, 2) The overall amplitude of the modified Regge-Wheeler potential is enhanced (i.e. experiences a positive-definite shift, as above). The size of all three effects is controlled by $\hat\beta$.
    \item We derive forecasted constraints on $\hat\beta$ and the associated shifts in the quasinormal frequency spectrum for upcoming ringdown observations. A single LISA or TianQin event can constrain these shifts at (up to) the $\mathcal{O}(\red{10^{-4}})$ level, with the potential of improving this further by two orders of magnitude by stacking multiple events. In the regime of small shifts/small $\hat\beta$ we investigated, these shifts linearly depend on $\hat\beta$ and hence this parameter is constrainable at (up to) the $\mathcal{O}(\red{10^{-4}})$ level (plus potential improvements from stacking). A suitably loud LVK observations, on the other hand, can provide analogous constraints at (up to) the $\mathcal{O}(10^{-2})$ level. Table \ref{tab-SNRringdown} collects equivalent order-of-magnitude achievable constraints for several current and future detectors. 
\end{itemize}

Several avenues therefore suggest themselves for future work extending the above: Most importantly, it will be interesting to see what complementary bounds the even sector will place and whether a fully stable corner of parameter space remains for such theories when considering both odd and even parity perturbations. Considering the dynamics of the theory \eqref{eq-theory} in detail in other regimes, e.g. its cosmological dynamics/limit as briefly discussed here, promises to further constrain the relevant parameter space. More generally speaking, we hope this paper provides a stepping stone towards identifying the landscape of which (if any) black hole solutions with time-dependent scalar hair are fully stable and hence are of particular interest for both theoretical and observational follow-up investigations.

\section*{Acknowledgments}
We thank Eugeny Babichev and Kazufumi Takahashi for useful discussions. SS is supported by an STFC studentship. JN is supported by an STFC Ernest Rutherford Fellowship (ST/S004572/1). In deriving the results of this paper, we have used xAct~\cite{xAct} and \cite{ringdown-calculations}. For the purpose of open access, the authors have applied a Creative Commons Attribution (CC BY) licence to any Author Accepted Manuscript version arising from this work.
\\

\noindent{\bf Data availability} Supporting research data are available on reasonable request from the authors.

\appendix
\section{Hairy black hole solutions}\label{app-hair}
Hairy black hole solutions in scalar-tensor theories can broadly be divided in two categories, depending on whether the scalar is static or contains a linear time-dependence. We here briefly review both, concentrating more on the latter case, which is the focus of the present paper. They each violate no-hair theorems in different ways. In the context of shift-symmetric theories ($\phi\rightarrow\phi+c$), \cite{Hui:2012qt} proved that black holes cannot support non-trivial scalar profiles if the following assumptions apply \footnote{A nice summary of the theorem and its consequences is provided in \cite{Lehebel:2018zga,Lecoeur:2024kwe}, including this itemised list.}
\begin{itemize}
    \item[1.] The spacetime is spherically symmetric, static and asymptotically flat,
    \item[2.] The scalar field is static $\phi(r)$ and has a vanishing derivative $\phi'$ at infinity,
    \item[3.] The norm of the current associated with the shift-symmetry is finite down to the horizon,
    \item[4.] The action contains a canonical kinetic term $X\subset G_2$,
    \item[5.] All $G_i(X)$ functions are analytical at $X=0$.
\end{itemize}

Time-dependent scalars automatically violate assumption 2, while static profiles will require the violation of an assumption other than 2. For reference, the hairy solution studied in the main text violates assumptions 2, 4 and 5.

\subsection{Static scalar}
We begin by considering solutions where the metric and the scalar are both static and spherically symmetric
\begin{align}
ds^2&=-B(r)dt^2+\frac{1}{B(r)}dr^2+r^2d\Omega^2, \nn \\
\bar\phi & =\phi(r).
\label{sss-backg}
\end{align}

Arguably the most common and minimal setup to achieve such kind of radial scalar hair is via a linear coupling to the Gauss-Bonnet invariant \cite{Sotiriou:2013qea,Sotiriou:2014pfa}, given by the choices
\begin{align}
    G_2&=\eta X, & G_4&=\zeta, & G_5&=\alpha \ln{|X|},
\end{align}
where $\eta$, $\zeta$ and $\alpha$ are constants and the other $G_i$'s are set to zero. The $G_5$ term is indeed equivalent to the linear coupling $\mathcal{L}\supseteq\alpha\phi\mathcal{G}$, where $\mathcal{G}$ is the Gauss-Bonnet invariant \cite{Kobayashi:2011nu}.

There exist, however, more general ways of breaking assumptions in no-hair theorems and thus constructing hairy solutions of this kind. \cite{Babichev:2017guv} showed that including at least one of the following
\begin{align}
    G_2&\supseteq\sqrt{X}, & G_3&\supseteq\ln{|X|}, & G_4&\supseteq\sqrt{X}, & G_5&\supseteq\ln{|X|},
    \label{r-hair}
\end{align}
to an overall shift-symmetric action is sufficient to potentially enact a radial-dependent scalar profile. These additional operators are in principle of notable interest as, unlike in the scalar Gauss-Bonnet case, they result in a current with a finite norm at the horizon \cite{Babichev:2017guv}. They do, however, come with potential issues, as it was shown that operators of the type \eqref{r-hair} other than the Gauss-Bonnet coupling lack a Lorentz invariant ($X=0$) solution in Minkowski spacetime \cite{Saravani:2019xwx}. If such solutions are desired, this requirement can therefore be used to eliminate all solutions in \eqref{r-hair}, except for the scalar-Gauss-Bonnet solution. Having said this, we note that the Lorentz-invariant $X=0$ vacuum is also not a stable solution in Galileon cosmologies consistent with CMB constraints \cite{Barreira:2013jma}, so the absence of a healthy $X=0$ solution need not imply the absence of relevant solutions in settings where Lorentz invariance is (already) spontaneously broken.

\subsection{Time-dependent scalar}
Let us now turn our attention to linearly time-dependent scalars, which directly violate assumption 2 and therefore require in principle no additional violations to generate hair. Table \ref{lit-rev} provides a comprehensive summary of the discussion presented here.

\subsubsection{Existence of solutions}
They were firstly found in \cite{Babichev:2013cya} for the theory given by
\begin{align}
    G_2&=-2\Lambda+2\eta X, & G_4&=\zeta+\beta X,
    \label{john}
\end{align}
with $\eta$, $\zeta$ and $\beta$ being constants. The $\beta$-term introduces a non-minimal coupling of the scalar field to the Einstein tensor (i.e. $\beta G^{\mu\nu}\phi_\mu\phi_\nu$)\footnote{To check this, note that the $\beta$-term contributes to $G_{4X}=\beta$. One can then use the Ricci identity of commuting covariant derivatives alongside integration by parts to show that
\begin{equation}
    G_{4X}\left[(\square\phi)^2-\phi^{\mu\nu}\phi_{\mu\nu}\right]=\beta R^{\mu\nu}\phi_\mu\phi_\nu.
\end{equation}
Combining this with $G_4R\ \ \backin\ \ -\beta\frac{1}{2}g^{\mu\nu}R\phi_\mu\phi_\nu$ gives us the coupling to the Einstein tensor.
} which has been referred to as the `John term' in \cite{Charmousis:2011bf} and plays an essential role in driving expansion. This result was then generalised in \cite{Kobayashi:2014eva} to the shift-symmetric ($\phi\rightarrow\phi+c$) and reflection-symmetric ($\phi\rightarrow-\phi)$ Horndeski subset, of which \eqref{john} is an example.\footnote{Other non-stealth black hole solutions were also found for this Horndeski subclass, but we will not be discussing them here.} In particular, it is the shift symmetry of the scalar which allows it to posses a linear time-dependence while keeping the background metric static, as the background equations of motion will only depend on derivatives of $\phi$ \cite{Babichev:2016rlq}.

As a specific case of \eqref{r-hair}, \cite{Babichev:2017guv} considered the theory given by
\begin{align}
    G_2&=\eta X, & G_4&=\zeta+\beta\sqrt{X}
\end{align}
and extended the construction of hairy black holes to the time-dependent background. The metric solution in this case was found to be non-stealth and hence we will not consider it here any further, but rather leave its stability analysis as a potential extension of this work.

It was also investigated in \cite{Babichev:2012re,Babichev:2016fbg} whether these time-dependent backgrounds could also be solutions of Horndeski theories incorporating $G_3$ such as the Cubic galileon, given by
\begin{align}
    G_2&=-2\Lambda\zeta+2\eta X, & G_3&=-2\gamma X & G_4&=\zeta.
    \label{cG}
\end{align}

This choice breaks the reflection symmetry of the previously studied theories but keeps the shift symmetry. However, no exact stealth black holes solutions of the form of \eqref{backg} were found. Instead, some approximated analytical expressions were given for different asymptotics, which we will not discuss here.

A breaking of shift-symmetry was also studied in \cite{Minamitsuji:2018vuw}, finding that solutions of the form \eqref{backg} cannot exist there. This is no longer true, however, in beyond Horndeski theories, where such background configurations were found to be solutions of shift-symmetric \cite{Motohashi:2019sen,Takahashi:2020hso} and shift-symmetry breaking quadratic DHOST \cite{BenAchour:2018dap}.

\subsubsection{Stability}
Despite being exact solutions to a considerably large class of scalar-tensor theories, backgrounds of the form of \eqref{backg} were quickly shown to be prone to instability issues. Odd parity perturbations were initially argued in \cite{Ogawa:2015pea} to possess either ghost or gradient instabilities close to the black hole horizon. It was later pointed out in \cite{Babichev:2018uiw} that this statement, made on the unboundedness of the Hamiltonian density, was coordinate-dependent and therefore not a good stability criterion, which should be independent of coordinates.\footnote{We thank Eugeny Babichev for related discussions.} In fact, it was shown there that stability could be attained in some time slicings for odd parity modes and even parity modes with $\ell=0$. The assumption of reflection symmetry was dropped from the stability analysis in \cite{Takahashi:2016dnv}, thus including also the non-stealth solutions found in \cite{Babichev:2016fbg}, and it was found that odd parity perturbations could be stable in some subclasses of shift-symmetric Horndeski. This analysis was then extended to the shift- and reflection-symmetric subclass of DHOST theories in \cite{Takahashi:2019oxz}, showing that odd parity perturbations there are stable and propagate with the same effective potential as in GR, i.e. the Regge-Wheeler potential.

However, higher-$\ell$ even modes for the same stealth solutions in DHOST were then investigated in \cite{deRham:2019gha}, finding that they were unfortunately plagued with instability or strong coupling issues. In particular, it was found that even modes coming from scalar perturbations were always everywhere unstable (c.f. to just close to the horizon in \cite{Ogawa:2015pea}). This result was later confirmed in \cite{Takahashi:2021bml}, which also found even gravitational modes to suffer from instabilities or be strongly coupled.
This then leaves us with no known well-behaved stealth black hole solutions with a linearly time-dependent scalar. Being $X=\textit{const}$ a key requirement for the results of \cite{deRham:2019gha} to hold, one can hope even parity perturbations might be stable in this configuration. We have shown in this paper that there exists an exact hairy black hole solution with non-constant X for which odd parity perturbations are stable.

In the context of non-exact stealth black hole solutions, which as mentioned before is not the focus of this paper, \cite{Minamitsuji:2019tet} looked at static solutions with a non-constant X and their stability against odd parity perturbations, finding that some configurations can be stable. When including time-dependence, however, no stable configuration was found.

\section{Background equations of motion}\label{app-EOMs}
Here we explore in more detail the background equations of motion. Writing the Lagrangian \eqref{root-X-action} as
\begin{equation}
    \mathcal{L}=\mathcal{L}_{GR}+\mathcal{L}_{\eta}+\mathcal{L}_{\lambda},
    \label{lagr}
\end{equation}
with
\begin{align}
    \mathcal{L}_{GR}&=R-2\Lambda,\nn\\
    \mathcal{L}_{\eta}&=2\eta\sqrt{X},\nn\\
    \mathcal{L}_{\lambda}&=\lambda\left(R\sqrt{X}+\frac{1}{2\sqrt{X}}\left((\square\phi)^2-(\phi_{\mu\nu})^2\right)\right),
\end{align}
the equations of motion from for the metric tensor and the scalar can respectively be written as
\begin{align}
    \mathcal{E}_{\mu\nu}&=\mathcal{E}_{\mu\nu}^{GR}+\mathcal{E}_{\mu\nu}^\eta+\mathcal{E}_{\mu\nu}^\lambda=0,\label{gEOM}\\
    \mathcal{E}_\phi&=\mathcal{E}_\phi^\eta+\mathcal{E}_\phi^\lambda=0\label{phiEOM},
\end{align}
where we introduce the notation $\mathcal{E}_{\mu\nu}^i=\frac{\delta S_i}{\delta g^{\mu\nu}}$, $\mathcal{E}_\phi^i=\frac{\delta S_i}{\delta \phi}$, and we have used the fact that $\mathcal{E}_\phi^{GR}=0$ since it is $\phi$-independent. Full expressions for $\{\mathcal{E}_{\mu\nu}^{GR},\mathcal{E}_{\mu\nu}^\eta,\mathcal{E}_{\mu\nu}^\lambda,\mathcal{E}_\phi^{GR},\mathcal{E}_\phi^\eta,\mathcal{E}_\phi^\lambda\}$ are given below.
We note that the above equations of motion satisfy an interesting relation given by
\begin{align}
    &\nabla^\mu\mathcal{E}_{\mu\nu}=-\phi_\nu\mathcal{E}_{\phi}.
    \label{constraint}
\end{align}
This is in fact satisfied independently by the three different lagrangians in \eqref{lagr}. One can easily check from the expressions \eqref{etag} and \eqref{etaphi} that $\nabla^\mu\mathcal{E}_{\mu\nu}^\eta=-\phi_\nu\mathcal{E}_{\phi}^\eta$. On the other hand, showing that $\nabla^\mu\mathcal{E}_{\mu\nu}^\lambda=-\phi_\nu\mathcal{E}_{\phi}^\lambda$ is more cumbersome and is therefore shown in \cite{ringdown-calculations} for the interested reader. Then, one is left with $\nabla^\mu\mathcal{E}_{\mu\nu}^{GR}=-\phi_\nu\mathcal{E}_{\phi}^{GR}$. As was argued before, $\mathcal{E}_{\phi}^{GR}=0$ due to the lack of $\phi$-dependence, which in turn reassures that Bianchi identities are satisfied, i.e. $\nabla^\mu\mathcal{E}_{\mu\nu}^{GR}=0$.

In this paper we are focusing on so-called `stealth' solutions, where despite of dynamical modifications induced by the scalar field, the metric background solution is as it would be in standard GR (in our case: a standard SdS solution) and hence $\mathcal{E}_{\mu\nu}^{GR}=0$. In that case equations \eqref{gEOM} and \eqref{phiEOM} imply that this particular background solution with a non-trivial scalar profile satisfies both
\begin{align}
    \label{scalarEOMcond}
    &\mathcal{E}_{\mu\nu}^\eta=-\mathcal{E}_{\mu\nu}^\lambda,\\
    &\mathcal{E}_{\phi}^\eta=-\mathcal{E}_{\phi}^\lambda.
    \label{metricEOMcond}
\end{align}
Here we write in full the following covariant expressions for the terms
 $\{\mathcal{E}_{\mu\nu}^{GR},\mathcal{E}_{\mu\nu}^\eta,\mathcal{E}_{\mu\nu}^\lambda,\mathcal{E}_\phi^{GR},\mathcal{E}_\phi^\eta,\mathcal{E}_\phi^\lambda\}$

\begin{widetext}
    \begin{align}
    \mathcal{E}_{\mu\nu}^{GR}&=R_{\mu\nu}-\frac{1}{2}g_{\mu\nu}\left(R-2\Lambda\right),\\
    \mathcal{E}_{\mu\nu}^{\eta}&=-\frac{\eta}{\sqrt{X}}\left(Xg_{\mu\nu}+\frac{1}{2}\phi_\mu\phi_\nu\right),\label{etag}\\
    \mathcal{E}_{\mu\nu}^{\lambda}&=\frac{\lambda}{2\sqrt{X}}\Bigg[-2X\left(R_\mn-\frac{1}{2}g_\mn R\right)+\frac{1}{2}\phi_\mu\phi_\nu\left(R-\frac{1}{2}(\square\phi)^2\right)+\frac{1}{2}g_\mn\left(2R_{\sigma\rho}\phi^\sigma\phi^\rho+\phi_{\sigma\rho}\phi^{\sigma\rho}-(\square\phi)^2\right)+\phi_\mn\square\phi\nn\label{lambdag}\\
    &\hspace{60pt}-2\phi^\sigma\left(\phi_{\{ \mu}R_{\nu\} \sigma}+\phi_{\{\mu}\phi_{\nu\}\sigma}\right)-\phi_{\nu\sigma}\phi_\mu^\sigma-R_{\mu\sigma\nu\rho}\phi^\sigma\phi^\rho\nn\\
    &\hspace{60pt}+\frac{1}{2X}\Big(2\phi^\sigma\phi_{\sigma\rho}\phi_{\{\mu}\phi_{\nu\}}^\rho-\phi_\mn\phi^\sigma\phi_{\rho\sigma}\phi^\rho-\frac{1}{2}\phi_\mu\phi_\nu\phi_{\sigma\rho}\phi^{\sigma\rho}+\phi_{\mu\sigma}\phi^\sigma\phi_{\nu\rho}\phi^\rho\nn\\
    &\hspace{100pt}+g_\mn\phi^\sigma\phi^\rho\left(\square\phi\phi_{\sigma\rho}-\phi_{\rho\gamma}-\phi_{\rho\gamma}\phi_\sigma^\gamma\right)\Big)\Bigg],\\
    \mathcal{E}_{\phi}^{GR}&=0,\\
    \mathcal{E}_{\phi}^{\eta}&=\frac{\eta}{\sqrt{X}}\left(\square\phi+\frac{1}{2X}\phi^\mu\phi^\nu\phi_{\mu\nu}\right),\label{etaphi}\\
    \mathcal{E}_{\phi}^{\lambda}&=\frac{\lambda}{\sqrt{X}}\Bigg[\frac{1}{2}R\square\phi-2\phi_\mn\phi^\mn+\square^2\phi-X\square\phi-R_\mn\phi^\mu\phi^\nu-\frac{3}{2}\phi_\mn\phi^\mu\phi^\nu\nn\\
    &\hspace{50pt}+\frac{1}{2X}\left(-3\phi^\mu\phi^\nu\phi_{\mu\sigma}\phi_\nu^\sigma+\phi_{\sigma\nu}\left(-\phi^\mn\phi_\mu^\sigma+\frac{1}{2}\square\phi\phi^{\sigma\nu}\right)+\phi^\mu\left(\frac{1}{2}R\phi^\nu\phi_\mn-2\phi_\mu^\nu\phi^\rho R_{\rho\nu}-\phi^\rho\phi_\nu^\sigma R^\nu_{\rho\sigma\mu}\right)\right)\nn\\
    &\hspace{50pt}+\frac{3}{(2X)^2}\phi^\mu\phi^\nu\phi_{\sigma\rho}\left(-\phi_\mu^\sigma\phi_\nu^\rho+\frac{1}{2}\phi_\mn\phi^{\sigma\rho}\right)\Bigg],
    \label{EOMs}
\end{align}
\end{widetext}
where we have used again the standard definition of symmetric and antisymmetric tensors \eqref{sym-tensors}.

As a consequence of the shift-symmetry of the theory considered here, the scalar equation of motion can be rewritten as a conservation equation
\begin{align}
    \mathcal{E}_{\phi}&=-\nabla_\mu J^\mu=0,
\end{align}
where the current $J^\mu$ can similarly be written as
\begin{align}
    J^\mu&=J^\mu_\eta+J^\mu_\lambda
\end{align}
with \cite{Sotiriou:2014pfa}
\begin{align}
    J^\mu_\eta&=-\frac{\eta}{\sqrt{X}}\phi^\mu,\\
    J^\mu_\lambda&=\frac{\lambda}{2\sqrt{X}}\Bigg[G^{\mn}\phi_\nu-\frac{1}{2X}\Big(\phi^\mu\left(\left(\phi_{\nu\sigma}\right)^2-\left(\square \phi\right)^2\right)\nn\\
    &\hspace{80pt}+2\phi^{\mu\nu}\left(\square\phi\phi_\nu-\phi_{\nu\sigma}\phi^\sigma\right)\Big)\Bigg].
\end{align}
Note that $\mathcal{E}_{\phi}^\eta=-\nabla_\mu J^\mu_\eta$ and $\mathcal{E}_{\phi}^\lambda=-\nabla_\mu J^\mu_\lambda$. The norm of the current is given by
\begin{align}
    J^2\equiv &J_\mu J^\mu=\frac{1}{6r(\lambda+\eta r^2)}\Bigg[384M\eta^2-r\Big(27\lambda^3\Lambda^2+108\eta^3 r^2\nn\\
    &+9\eta\lambda^2\Lambda(20+3\Lambda r^2)+4\eta^2\lambda(75+29\Lambda r^2)\Big)\Bigg].
\end{align}
Hence, the norm of the current is found to be regular for all $r$ down to the black hole horizon.

\section{$\Lambda$ effects on QNMs}\label{large-r}
Here we explore the effect that a cosmological constant term $\Lambda$ has on the emission of quasinormal modes, and use this as a complementary check of the semi-analytical WKB method used in the main text. One would expect that such an effect is strongly suppressed for two reasons: 1) the small value of the cosmological constant, and 2) the fact that quasinormal mode emission should be governed by the local dynamics around the source, and $\Lambda$ contributes as a large-$r$ term to the spacetime metric
\begin{equation}
    B=1-\frac{2M}{r}-\frac{1}{3}\Lambda r^2.
\end{equation}

Whether this really impacts at large $r$ or not is of course dependent on the value of $\Lambda$. As was shown in \cite{Barausse:2014tra}, for cosmological values of $\Lambda$ (i.e. $\Lambda/\mpl^2=10^{-52}\text{m}^{-2}$ in natural units) its effect on the QNMs is negligible. For example, for a black hole of $10^6M_\odot$, the correction to the quasinormal modes appears at the $10^{-34}$ level.

The SdS solution also allows for values of $\Lambda$ larger than the one for the cosmological constant. In particular, the range of values allowed for $\Lambda$ which preserve the 2 horizon nature of SdS is $0\leq\Lambda\leq\frac{1}{9M^2}$. QNM values spanning this full range of $\Lambda$ were calculated in \cite{Zhidenko:2003wq}. In the cases considered there, where the black hole and cosmological horizon are of comparable size, it is not surprising that deviations in the frequencies become quite significant.

We employ here the WKB method to obtain quasinormal frequencies in the two same ways as has been done in the main text. First, we apply the package \cite{wkb-package} directly and obtain numerical solutions for specific values of $\Lambda$. By doing so, we find that our results agree well with \cite{Zhidenko:2003wq}. Secondly, we use the $6^{th}$ order WKB formula \eqref{wkb-formula} and take derivatives of the potential while keeping $\Lambda$ unspecified. Hence, we are able to obtain an expression for $\omega(\Lambda)$. As was the case with $\omega(\beta)$ in the main text, the expression is cumbersome and therefore unfeasible to write here. One can find such expression in \cite{ringdown-calculations}. Here, we instead Taylor expand around the zero value for $\Lambda$ and obtain\footnote{Higher order contributions can be obtained in \cite{ringdown-calculations}.}
\begin{align}
    M\omega=M\omega_0&-[1.67631-0.33463 i]\cdot M^2\Lambda\nn\\
    &-[3.85847-0.89970 i]\cdot M^4\Lambda^2\nn\\
    &-[17.4684-6.04082 i]\cdot M^6\Lambda^3+\mathcal{O}(\Lambda^4).
    \label{QNM-Lambda}
\end{align}
where we recover $M\omega_0$ as the Schwarzschild quasinormal frequencies. Using this semi-analytical expressions, we are also able to recover the same results as in \cite{Zhidenko:2003wq}, gaining precision for smaller $\Lambda$. This provides a non-trivial validation test of the semi-analytical prescription, which we developed in the main text to constrain $\hat{\beta}$.\footnote{The maximum of the potential is shifted due to $\Lambda$. Employing the \textit{light-ring expansion} \cite{Franciolini:2018uyq}, for small values of $\Lambda$, this is given by
\begin{align}
    r_*^{\text{max}}
    &=3.28\cdot M - 2.85\cdot M^3\Lambda+\mathcal{O}(\Lambda^2),
\end{align}
}
Finally, for a black hole of $10^6M_\odot\simeq10^9\text{m}$\footnote{Note the the mass of the Sun in geometric units is around $1.5\text{km}\simeq10^3\text{m}$.} and a cosmological constant of $\Lambda=10^{-52}\text{m}^{-2}$, we can easily use equation \eqref{QNM-Lambda} to confirm that the first correction to the quasinormal frequencies appears at the $10^{-34}$ level, in agreement with \cite{Barausse:2014tra}.

\bibliographystyle{apsrev4-1}
\bibliography{QNM-SdS-t-scalar}
\end{document}